\documentclass[onecolumn,amsmath,nofootinbib,secnumarabic,eqsecnum,amssymb,12pt]{revtex4}
\pdfoutput=1

\usepackage{bm}

\usepackage{epsf}
\usepackage{graphicx,epsfig}
\usepackage{amsfonts}
\usepackage{amssymb}
\usepackage{xcolor}
\usepackage{subfigure}

\definecolor{mine}{rgb}{0.2,0.1,0.7}
\definecolor{bb}{rgb}{0.3, 0.5, 1}
\definecolor{bg}{rgb}{0.1, 0.1, 0.5}
\usepackage[%
    pdftitle={The title of your letter},
    pdfauthor={Your name},
    pdfsubject={The subject of the letter},
    pdfkeywords={Any keywords},
     colorlinks=true,  
    linkcolor=mine,
     citecolor=bg, 
     urlcolor=bg, 
     pdfnewwindow=true,
     pagebackref=true,
     filecolor=bb,
     pdffitwindow=false,     
     pdfstartview={FitH}    
]{hyperref}

\def\CO{{\cal O}}
\def\half{\frac12}
\def\L{{\cal L}}
\def\a{A}
\def\l{\lambda}
\def\s{\sigma}
\def\A{{\cal A}}

\def\H{H}
\def\g{\gamma}
\def\c{c_{{\cal D}}}

\def\invc{\left( \frac{1}{\c^2}-1\right)}
\def\b{\frac{1}{f}}
\def\e{M^2}
\def\M{M_P}
\def\hP{\Pi}
\def\pQ{\partial Q}
\def\dQ{\dot Q}
\def\pdQ{\partial \dot Q}

\def\sd{\dot \sigma}

\def\matter{{\rm brane}}

\def\tN{{\tilde N}}
\def\th{\tilde{h}}

\def\tB{{\tilde \beta	}}
\def\tR{{\tilde R}}
\def\tE{{\tilde E}}
\def\tpone{     { \alpha}^{(1)}     }
\def\tptwo{     { \alpha}^{(2)}     }
\def\tBone{\tB^{(1)}}

\def\E{E}

\def\p{\phi}
\def\PX{P_{,X}}

\def\z{\zeta}

\def\deltaN{\delta N}

\def\X{\kappa}

\def\cs{c_s}

\def\teps{{\tilde \epsilon}}
\def\tH{{\tilde H}}

\def\w{w}

\def\po{p}

\newcommand{\dn}[2]{{\mathrm{d}^{{#1}}{{#2}}}}

\def\rc{\rho_c}

\def\q{\alpha}

\def\ca{c_\sigma}
\def\ce{c_s}
\def\cae{c_{\s s}^2}

\def\sdd{\ddot \sigma}

\def\dQsi{\dot Q_{\sigma}}

\def\Qsi{Q_{\sigma}}
\def\pQsi{\partial Q_{\sigma}}
\def\LapQsi{\partial^2 Q_{\sigma}}
\def\pdQsi{\partial \dot Q_{\sigma}} 

\def\dQs{\dot Q_{s}}
\def\Qs{Q_{s}}
\def\pQs{\partial Q_{s}}
\def\LapQs{\partial^2 Q_{s}}
 
\def\kineticperp{ \dQs^2  }
\def\gradperp{ (\pQs)^2}
\def\doneLaponeperp{  \dQs \LapQs  }
\def\kineticgradperp{   \pQs \dQs   }
\def\doneperppuipuj{  \dQs \partial^i \partial^j \Qs  } 
\def\pdipdjQsi{  \partial_i \partial_j \Qsi  }

\def\poneLaponeperp{  \pQs \LapQs  }

\def\N{N}

\def\eps{\epsilon}

\newcommand\bk{\boldsymbol{k}}
\def\Kone{K_1}
\def\Ao{S_1}
\def\At{S_2}

\def\tM{{\tilde M}}

\def\R{\zeta}

\def\T{T_{\sigma s}}

\def\Sa{S^{(\s \s \s)}}
\def\Ss{S^{(\s ss)}}

\def\tg{{\tilde g}}

\def\D{{\cal D}}

\def\tt{{\tilde t}}

\newcommand{\bea}{\begin{eqnarray}}
\newcommand{\eea}{\end{eqnarray}}
\newcommand\be{\begin{equation}}
\newcommand\ee{\end{equation}}
\newcommand\beq{\begin{equation}}
\newcommand\eeq{\end{equation}}

\newcommand{\nn}{\nonumber}

\newcommand{\refeq}[1]{(\ref{#1})}

\def\Tdot#1{{{#1}^{\hbox{.}}}}

\def\cg{c_{t}}
\def\qg{q}

\makeatletter
\renewcommand\section{\@startsection {section}{1}{\z@}%
                                 {-3.5ex \@plus -1ex \@minus -.2ex}
                                   {2.3ex \@plus.2ex}%
                                   {\normalfont\large\bfseries}}
\renewcommand\subsection{\@startsection{subsection}{2}{\z@}%
                                   {-3.25ex\@plus -1ex \@minus -.2ex}%
                                     {1.5ex \@plus .2ex}%
                                     {\normalfont\bfseries}}
\renewcommand\subsubsection{\@startsection{subsubsection}{3}{\z@}%
                                   {-3.25ex\@plus -1ex \@minus -.2ex}%
                                     {1.5ex \@plus .2ex}%
                                     {\normalfont\bf\itshape}}
                                    
\makeatother

\begin{document}

\begin{center}
{\Large \bf  
Primordial fluctuations and non-Gaussianities\\
\vspace{0.5cm}
from multifield DBI Galileon inflation}

\vskip 1.3cm
\centerline{\large S\'ebastien Renaux-Petel$^{1}$
, Shuntaro Mizuno$^{2}$ and Kazuya Koyama$^{2}$}

\vskip 0.8cm
{\em    ${}^1$ Centre for Theoretical Cosmology,\\
Department of Applied Mathematics and Theoretical Physics,\\
University of Cambridge, Cambridge CB3 0WA, UK    \\[0.3in]}
{\em    ${}^2$  Institute of Cosmology and Gravitation,\\
University of Portsmouth, Portsmouth PO1 3FX, UK \\[0.3in]}

\end{center}

\begin{center}
{\bf
Abstract}
\end{center}

We study a cosmological scenario in which the DBI action governing the motion of a D3-brane in a higher-dimensional spacetime is supplemented with an induced gravity term. The latter reduces to the quartic Galileon Lagrangian when the motion of the brane is non-relativistic and we show that it tends to violate the null energy condition and to render cosmological fluctuations ghosts. There nonetheless exists an interesting parameter space in which a stable phase of quasi-exponential expansion can be achieved while the induced gravity leaves non trivial imprints. We derive the exact second-order action governing the dynamics of linear perturbations and we show that it can be simply understood through a bimetric perspective. In the relativistic regime, we also calculate the dominant contribution to the primordial bispectrum and demonstrate that large non-Gaussianities of orthogonal shape can be generated, for the first time in a concrete model. More generally, we find that the sign and the shape of the bispectrum offer powerful diagnostics of the precise strength of the induced gravity.

\noindent

\vfill

\newpage
\setcounter{page}{1}

\tableofcontents

\newpage

\section{I\lowercase{ntroduction}}

On-going measurements of the cosmic microwave background (CMB) anisotropies by the Planck satellite \cite{Planck} promise to provide us with yet more precise information on the very early universe than those we gained from the WMAP satellite \cite{WMAP}. In this respect, while Gaussianity of the primordial fluctuations is up until now still preferred by the data \cite{Fergusson:2010dm,Komatsu:2010fb}, a tremendous amount of attention is currently paid to their possible deviation from perfect Gaussian statistics (see \cite{Komatsu:2010hc,Liguori:2010hx} for recent observational reviews). Indeed, any such detection would shed some light on the interactions of the field(s) driving inflation or its alternatives, a piece of information that can not be probed by the Gaussian linear theory. These interactions are very small in slow-roll single field inflationary models, which therefore predict a very small amplitude of primordial non-Gaussianities \cite{Maldacena:2002vr}. However, going beyond these simplest scenarios have demonstrated that sizeable non-Gaussianities can be generated in a wide variety of early universe models (see for instance the reviews \cite{Chen:2010xka, Koyama:2010xj, Wands:2010af,Byrnes:2010em} and references there in).

Amongst them, the Dirac-Born-Infeld (DBI) model \cite{Silverstein:2003hf,Alishahiha:2004eh,Chen:2004gc,Chen:2005ad,Chen:2005fe}, in which the inflaton is identified with the radial coordinate of a probe D3-brane in a warped geometry, has attracted considerable amount of attention as an inflationary scenario inspired by string theory, producing large non-Gaussianities, and possibly alleviating the famous eta-problem \cite{copeland-1994-49,Stewart:1994ts}. It is also one of the few $k$-inflationary models \cite{ArmendarizPicon:1999rj,Garriga:1999vw} whose higher derivative structure is known not to be spoiled by quantum corrections \cite{Kabat:1999yq}. One of the lessons we learnt from the extensive studies to put constraints on DBI inflation \cite{Shandera:2006ax,Kecskemeti:2006cg,Lidsey:2006ia,Baumann:2006cd,Bean:2007hc,Lidsey:2007gq,
Peiris:2007gz,Kobayashi:2007hm,Lorenz:2007ze,Bean:2007eh,Bird:2009pq}
is that so-called ultraviolet models are under strain from observations because they generate a too high level of non-Gaussianities\footnote{Concerns also exist about their consistent embedding in string theory \cite{Baumann:2006cd,Bean:2007eh,Chen:2008hz}.}. 
However, this applies only to single field models of DBI inflation, in which the angular degrees of freedom of the brane are assumed to be frozen. As was shown in Ref. \cite{Langlois:2008wt} and in further details in \cite{Langlois:2008qf}, relaxing this assumption and allowing the brane to evolve and fluctuate in the angular directions (see e.g \cite{Kehagias:1999vr,Steer:2002ng,Brax:2002jj,Easson:2007fz,Agarwal:2011wm}) alleviates these constraints. Indeed, 
in such a multiple field scenario,
the amplitude of equilateral non-Gaussianities is suppressed as a result of the transfer from entropic to curvature perturbations
(see \cite{Arroja:2008yy,RenauxPetel:2008gi,Langlois:2009ej,Mizuno:2009cv,RenauxPetel:2009sj,Gao:2009at,Mizuno:2009mv,Gao:2009gd,Cai:2009hw,Pi:2011tv} for subsequent studies on multi-field inflationary models with non-canonical kinetic terms).

Recently, de Rham and Tolley proposed a very interesting extension of the DBI action \cite{deRham:2010eu} which generalizes and unifies single field DBI and the modified gravity model called the Galileon \cite{Nicolis:2008in, Deffayet:2009wt,Deffayet:2009mn} (see the work by Horndeski \cite{Hordenski} for an early study of a related more general class of models and Refs.~\cite{Chow:2009fm,Silva:2009km,Kobayashi:2009wr,Kobayashi:2010wa,
Gannouji:2010au,DeFelice:2010jn,DeFelice:2010gb,Creminelli:2010ba,DeFelice:2010pv,Ali:2010gr,DeFelice:2010nf,Burrage:2010cu,Mota:2010bs,Nesseris:2010pc,Creminelli:2010qf,DeFelice:2010as,Khoury:2011da,Levasseur:2011mw,Liu:2011ns,Brax:2011sv,deRham:2011by} for some studies of cosmology based on the Galileon field). Following the geometrical picture of a D3-brane embedded in a five-dimensional spacetime (see also \cite{Goon:2011qf, Goon:2011uw, Burrage:2011bt,Trodden:2011xh}), they showed that only a finite number of terms are allowed that give rise to second-order equations of motion for the brane position modulus, and that these terms reduce to the so-called Galileon Lagrangians in the non-relativistic limit in which the motion of the brane is slow (see \cite{Mizuno:2010ag} for a study of the non-Gaussianities generated by one of these terms). This relativistic extension of the Galileon model has later been generalized to the case of an arbitrary number of extra dimensions in Ref. \cite{Hinterbichler:2010xn}, based on the work \cite{Charmousis:2005ey} that studied matching conditions for distributional sources of arbitrary co-dimension in the context of Lovelock gravity (see Refs.~\cite{Deffayet:2010qz, Pujolas:2011he, Kimura:2010di,Kobayashi:2010cm, Kamada:2010qe, DeFelice:2011zh, Kobayashi:2011pc,Deffayet:2011gz,Charmousis:2011bf,Kobayashi:2011nu,Gao:2011qe,DeFelice:2011uc} for other interesting 
generalizations in the single field case and their cosmological implications). Multifield extensions of the Galileon terms were extensively studied in Refs.~\cite{Andrews:2010km,Deffayet:2010zh,Padilla:2010de, Padilla:2010ir, Padilla:2010tj,Zhou:2010di} but the cosmology of the multifield DBI Galileon model started to be explored by one of us only very recently \cite{RenauxPetel:2011dv}.
In this paper, we give further details on this model formulated in an ambient ten-dimensional spacetime. In this case -- as relevant to string theory -- only an induced gravity term can be added to the DBI action \cite{Charmousis:2005ey,Hinterbichler:2010xn}. As we will see, this geometrical structure implies that it is useful to think of this model not as a modification of multifield DBI inflation, but rather as a multi scalar-tensor modification of gravity with a peculiar bimetric structure.

The layout of this paper is the following. In section \ref{Set-up} we introduce our model and derive the conditions under which it can sustain a phase of quasi de-Sitter expansion. We study the scalar and tensor linear perturbations about a cosmological background in section \ref{Lin} while section \ref{NGs} is devoted to the study of the non-Gaussianities of the primordial curvature perturbation. We give our conclusions in section \ref{Conclusion} and leave some technical details to several appendices.

\section{S\lowercase{et-up and homogneous evolution}}
\label{Set-up}

\subsection{Action and equations of motion}

The geometrical set-up we consider is the same as in standard multifield DBI inflation \cite{Langlois:2008qf}: we consider a D3-brane with tension $T_3$ evolving in a 10-dimensional geometry described by the metric
\be
ds^2 = h^{-1/2}(y^K)\,g_{\mu \nu}dx^\mu dx^\nu + h^{1/2}(y^K)\, G_{IJ}(y^K)\, dy^I dy^J \equiv H_{AB} dY^A dY^B
\ee
with coordinates $Y^A=\left\{x^\mu, y^I\right\}$, 
where $\mu=0,\ldots 3$ and $I=1,\ldots, 6$ (the label $I$ has been chosen in this way as below it will label the multiple effective scalar fields). An important role will be played by the induced metric on the 3-brane,
\be
\gamma_{\mu \nu} =H_{AB} \partial_\mu Y_{\rm (b)}^A \partial_\nu Y_{\rm (b)}^B 
\ee
where the  brane embedding is defined by the functions  $Y_{\rm (b)}^A(x^\mu)$, with the $x^\mu$ being the spacetime coordinates on the brane. In our case, they coincide with the first four bulk coordinates, so that, on writing
$Y_{\rm (b)}^A = (x^\mu, \varphi^I(x^{\mu}))$, we find
\be
\gamma_{\mu \nu} = h^{-1/2} \left( g_{\mu \nu}  + h \, G_{IJ} \partial_\mu \varphi^I \partial_\nu \varphi^J \right)
\label{induced-metric}\,.
\ee
As discussed in the introduction, our starting point four-dimensional effective action will be the sum of three terms:
\begin{itemize}
\item the DBI action, whose kinetic part, proportional to the world-volume encompassed by the brane\footnote{Strictly speaking, the DBI action also depends on bulk fields as well as on the gauge field living on the brane. Their cosmological consequences were analysed in Ref. \cite{Langlois:2009ej}.}, is simply a cosmological constant in terms of the induced metric:  $-T_3 \sqrt{-\gamma}$.
\item the Einstein-Hilbert Lagrangian associated to the induced metric: $\sqrt{-\g}R[\g]$. This term is absent in standard DBI inflation and will be the focus of our attention\footnote{Such an induced gravity term is present in bosonic string theory \cite{Corley:2001hg}. It is absent at tree level in type II superstring theories \cite{Fotopoulos:2002wy} but may be induced at 1-loop string level. We thank Liam McAllister for useful discussions on this point.}. 
\item the standard Einstein-Hilbert Lagrangian associated to the cosmological metric: $\sqrt{-g}R[g]$. 
\end{itemize}
The total action thus reads
\be
S= \int {\rm d}^4 x \left[ \frac{\M^2}{2}  \sqrt{-g} R[g]  +\frac{\e}{2} \sqrt{-\g} R[\g] + \sqrt{-g}  {\cal L}_\matter    \right]  
\label{action}
\ee
where $\M$ and $M$ are constant mass scales\footnote{If some matter was to be added to the action, we have in mind that it would be minimally coupled to the metric $g_{\mu \nu}$. Hence, we conventionally refer to the mass scale associated to the standard Einstein-Hilbert Lagrangian as $\M$. However, we should bear in mind that, because of the induced gravity, this is a priori not directly related to the Newton constant as would be measured in laboratory experiments.} and
\be
 {\cal L}_\matter= -\frac{1}{f(\phi^I)}\left(\sqrt{{\cal D}}-1\right) -V(\phi^I)\,.
 \label{brane-action}
\ee
Here, following the literature on brane inflation, we have introduced the rescaled variables
\be
f= \frac{h}{T_3} \; , \qquad \phi^I = \sqrt{T_3}\varphi^I\,,
\label{redef}
\ee
we have included potential terms in the brane action and we have defined
\beq
{\cal D} \equiv  \det(\delta^{\mu}_{\nu}+f \, G_{IJ} g^{\mu \rho}  \partial_{\rho} \p^I \partial_{\nu} \p^J )\,
\label{Ddef}
\ee
where $G_{IJ}(\phi^K)$ will play the role of a metric in the space of the scalar fields $\phi^I$. On defining the mixed kinetic terms for the scalar fields 
\be
X^{IJ} \equiv -\frac12 g^{\mu \nu}\partial_{\mu} \phi^I \partial_{\mu} \phi^J\,,
\label{def-XIJ}
\ee
it has been shown in \cite{Langlois:2008qf} that the explicit expression of ${\cal D}$ reads
\beq
{\cal D}=1-2f G_{IJ}X^{IJ}+4f^2 X^{[I}_IX_J^{J]} -8f^3 X^{[I}_IX_J^{J} X_K^{K]}+16f^4 X^{[I}_IX_J^{J} X_K^{K}X_L^{L]},
\label{def_explicit}
\eeq
where the brackets denote antisymmetrisation on the field indices and $X_{I}^{J}=G_{IK} X^{KJ}$. Similarly, one can express $ \sqrt{-\g}R[\g]$ in terms of the fields and the geometrical quantities associated to the cosmological metric, leading to a multifield relativistic extension of the quartic Galileon Lagrangian in curved spacetime. The resulting expression, that can be found in appendix \refeq{R[q]explicit}, is very intricate and rather obscures the physics which is at play by breaking the symmetry between the cosmological and the induced metric in the ``gravitational'' part of the action \refeq{action}. In the following, we rather try to keep this symmetry manifest by treating the induced gravity and Einstein-Hilbert action on equal footing. \\

Following this bimetric perspective, one can write the gravitational equations of motion -- obtained by varying the action \refeq{action} with respect to $g_{\mu \nu}$ -- in the compact form
\bea
\label{metric-eoms}
 \M^2  G^{\mu \nu}[g]+\e  \frac{\sqrt{{\cal D}}}{h^{3/2}}G^{\mu \nu}[\g] =T^{\mu \nu}_\matter 
\eea
where we have used the fact that
\be
\sqrt{-\gamma}=h^{-1} \sqrt{-g} \sqrt{{\cal D}}\,,
\ee
we have defined
\be
T^{\mu \nu} \equiv -\frac{2}{\sqrt{-g}} \frac{\delta \left( \sqrt{-g} {\cal L}_\matter   \right)}{\delta g_{\mu \nu}}= - \frac{1}{f} \sqrt{\cal D} \tg^{\mu \nu} -g^{\mu \nu} \left( V-\b \right)
\ee
and $\tg^{\mu \nu}$ denotes the inverse of the metric 
\be
\tg_{\mu \nu} \equiv  g_{\mu \nu}  + f \, G_{IJ} \partial_\mu \phi^I \partial_\nu \phi^J
\label{induced-tg}
\ee
(the latter is simply $h^{1/2} \gamma_{\mu \nu}$). As for the equations of motion for the scalar fields, they read
\be
\label{scalar-eoms}
\left(\frac{\e}{h^{3/2}} G^{\mu \nu}[\g]+\b \tg^{\mu \nu}\right) 
 \left( \delta^I_J+2 f G_{JK} X^{IK}_{\tg}  \right) 
 \left( \Pi_{\mu \nu}^J+\frac{f^{,J}}{4f} \frac{\tg_{\mu \nu}}{f}   \right)  
 -\frac{G^{IJ}}{f \sqrt{{\cal D}}} \left( V_{,J}+\frac{f_{,J}}{f^2}  \right)=0\,.
\ee
Here, $X^{IJ}_{\tg} \equiv -\frac12 \tg^{\mu \nu}\partial_{\nu} \phi^I \partial_{\mu} \phi^J$ and
\be
\Pi^I_{\mu \nu}  \equiv \nabla_\mu \nabla_\nu \phi^I+{\hat \Gamma}^I_{AB} \nabla_\mu \phi^A  \nabla_{\nu} \phi^B
\ee
where $\nabla_{\mu}$ denotes the covariant derivative associated to $g_{\mu \nu}$ and ${\hat \Gamma}^I_{AB}$ are the Christoffel symbols associated to ${\hat G}_{IJ} \equiv f G_{IJ}$. Note that, because of the induced gravity, second-order derivatives both of the metric and of the scalar fields enter into the gravitational \refeq{metric-eoms} and scalar fields \refeq{scalar-eoms} equations of motion, a property whose consequences have recently been explored in a slightly different context \cite{Deffayet:2010qz,Pujolas:2011he}.\\

It turns out that the warping -- the fact that $h$ (and hence $f$) depends non-trivially on the scalar fields -- complicates the understanding of the role of the induced gravity in a non essential way. Hence, in the remaining of the main body of this paper, we consider a constant warp factor, in which case our starting point action simplifies to
\be
S_{f={\rm cst}}= \int {\rm d}^4 x \left[ \frac{\M^2}{2}  \sqrt{-g} R[g]  +\frac{\tM^2}{2} \sqrt{-\tg} R[\tg] + \sqrt{-g}  {\cal L}_\matter    \right]  
\label{action-f=cst}
\ee
where $\tM^2=M^2/\sqrt{h}$ is constant. For simplicity, we also refer to the metric \refeq{induced-tg} as the induced metric.
The consequences of a non-trivial warping are addressed in appendix \refeq{warping}.

\subsection{Background evolution}
\label{Background}

In this section, we specialize the above equations to homogeneous configurations of the fields in a flat Friedmann-Lema\^itre-Robertson-Walker (FLRW) background spacetime, of metric
\be
g_{\mu \nu} dx^{\mu} dx^{\nu}=-dt^2 +a^2(t) d \boldsymbol{x}^2\,.
\ee
Like in standard brane inflation, it is useful to
introduce the background value of $\cal D$ \refeq{def_explicit}: 
\be
\c^2 \equiv 1- f  \dot \sigma^2\,,
\ee where $\sd \equiv \sqrt{G_{IJ} \dot \phi^I   \dot \phi^J}$ plays the role of an effective collective velocity of the fields\footnote{In this paper, we use the symbol $\c^2$, and not the more conventional $c_s^2$, because, contrary to standard brane inflation, this does not coincide with the speed of sound for (scalar) perturbations.}. The value of $\c^2$ distinguishes the slow-roll regime where $\c^2 \simeq 1$ and the brane action \refeq{brane-action} takes a canonical form, from the so-called relativistic, or DBI, regime, where the brane almost saturates its speed limit and $\c^2  \ll 1$. In that case, the whole non-linear structure of  the action \refeq{brane-action} must be taken into account and large non-Gaussianities are expected. It is in this phenomenologically more interesting regime that we are particularly interested in.

To derive the background equations of motion, it is not necessary to use the explicit expressions in appendix \refeq{R[q]explicit} for the geometrical quantities associated to the induced metric, like $R[\tg]$. It is rather more useful to notice that, in the background, the induced metric \refeq{induced-tg} takes the form of a flat FLRW metric whose scale factor is the cosmological one, $a$, but whose cosmic time $\tt$ is such that 
\be
d \tt=\c dt\,.
\label{induced-time}
\ee Introducing the corresponding Hubble and ``deceleration'' parameters, given respectively by $\tH \equiv  \frac{1}{a} \frac{d a}{d \tt} = \frac{H}{\c}$ and 
 \be
 \teps  \equiv  -\frac{1}{\tH^2} \frac{d \tH}{d \tt} =   \eps + s\,,
 \label{teps}
 \ee
 where $\eps \equiv - \dot H/H^2$ and $s \equiv \dot{c}_{\D}/(H \c)$, it is then straightforward to derive the expressions of geometrical quantities associated to the induced metric from well known results. For example, from $R_{tt}[g]=-3 H^2 (1-\eps)$, we deduce that $R_{\tt \tt}[\tg]=-3 \tH^2(1-\teps)$, and hence, using the transformation law of tensors: $R_{tt}[\tg]=-3 \c^2 \tH^2 (1-\teps)=-3 H^2 (1-\eps-s)$. Following this method, the modified Friedmann equations are readily found to be
\bea
\label{Friedmann1}
 3 H^2 \left(\M^2 +\frac{\tM^2}{\c^3} \right)=\rho_\matter = V+ \frac{1}{f}\left( \frac{1}{\c}-1 \right)
\eea
\bea
\M^2 H^2 \eps +\frac{ \tM^2 H^2}{\c} \left( \eps +s+\frac32 \invc \right)= \frac{\sd^2}{2\c}\,.
\label{Hdot}
\eea
We do not reproduce here the background equations of motion for the fields because they take a complicated and not very illuminating form. They can be found in full generality in appendix \refeq{warping}. As noticed above, second-order derivatives of the fields enter into the gravitational equation of motions (and vice-versa). However, solving the associated system of linear equations is not required to understand interesting properties of the homogeneous evolution. In particular, the way we derived Eq. \refeq{Hdot} makes it clear that second-order derivatives of the fields enter into it only through the parameter $s$. Hence, in a quasi de-Sitter inflationary spacetime in which the time evolution of every quantity is slow with respect to that of the scale factor -- \textit{i.e.} $\dot X/ (HX) \ll 1$ -- $s$ is not expected to qualitatively alter the solutions of \refeq{Hdot} and thus can be neglected. Note also that the combinations of mass scales that enter into Eqs. \refeq{Friedmann1} and \refeq{Hdot} are respectively $\M^2 +\tM^2/\c^3$ and $\M^2 +\tM^2/\c$. Hence, because of the non-zero velocity of the fields -- $\c \neq 1$ -- the effect of the bimetric structure is not a simple rescaling of $\M$. A parameter that will prove to be useful to measure this effect is
\be
\X \equiv  \frac{ \M^2+\frac{\tM^2}{\c} }{ \M^2+\frac{\tM^2}{\c^3}  }\,
\label{def-kappa}
\ee
that is comprised between $\c^2$ and $1$. Combining Eqs. \refeq{Friedmann1} and \refeq{Hdot}, we find (neglecting $s$):
\be
\eps =\frac{3}{2 \X} \frac{1-\c^2}{(\c f V+1-\c)}-\frac{3}{2}\frac{\tM^2}{(\c \M^2+\tM^2)}\invc \,.
\label{eps-explicit}
\ee
From the second term, one deduces that, for generic values of $\tM$, $\eps$ acquires a large negative contribution when $\c^2 \ll1$, which tends to violate the null energy condition and to hamper inflation. Barring cancellations between the positive and negative contributions to $\eps$ in \refeq{eps-explicit}, one thus finds that achieving a phase of quasi de-Sitter expansion in the relativistic regime -- $\c^2 \ll 1$ -- requires, in addition to the usual condition to enter into the DBI regime 
\be
\c f V \gg 1
\label{condition1}\,,
\ee
that 
\be
\tM^2 \ll \c^3 \M^2
\label{condition2}
\ee
(and hence $\X \simeq 1)$. At this point, given this restrictive condition on the mass scale associated to the induced gravity, it seems hopeless to expect anything new from multifield DBI Galileon inflation compared to standard multifield DBI inflation. However, this guess is not correct. To see this, let us introduce the dimensionless quantity 
\be
\q \equiv \frac{ f H^2 \tM^2}{\c^2}\,.
\label{def-alpha}
\ee
Combining this definition with Eq. \refeq{Friedmann1}, one finds that, in our regime, $3\q \simeq  \frac{\tM^2}{\M^2 \c^3} (\c f V)$ and hence, being the product of a small parameter by a large one, is not necessarily small. The approximate equation deduced from \refeq{eps-explicit}:
\be
\eps \simeq  \frac{3}{2 \c f  V}  (1-3\q) 
\label{eps-approx} 
\ee
then shows that the induced gravity can have a non negligible effect on the background evolution. Note also that this contribution to $\eps$ is always positive in the regime where the theory is ghost-free, as will be demonstrated by the analysis of cosmological perturbations in the next section.

\section{L\lowercase{inear cosmological perturbations}}
\label{Lin}

In this section, we study the dynamics of linear perturbations about a homogenous cosmological solution. For this purpose, one can envisage several approaches. One is to perturb up to first order the equations of motion. This is not suitable for quantizing the fluctuations but this is rather straightforward and one easily deduces from this the fact the two Bardeen potentials differ. In other words, there exists an effective anisotropic stress in the multifield DBI Galileon model, whose consequences for the growth of perturbations and in the late universe would be interesting to study. This is shown in appendix \refeq{Linear}. The modern approach to cosmological perturbations during inflation, pioneered by Mukhanov and Chibisov \cite{Mukhanov:1981xt} and Sasaki \cite{Sasaki1986} and further developed by Maldacena \cite{malda}, is rather to use the ADM formalism \cite{Salopek:1990jq} to calculate efficiently the action at second order in the perturbations. From this action, one can easily derive the equations of motion for the linear perturbations but also determine the normalization of the vacuum quantum fluctuations and hence the amplitude of primordial cosmological perturbations. A priori, one can apply this strategy to the explicit form of the action given in Eq. \refeq{explicit-action}. However, this expression is complicated and the resulting calculations are very cumbersome. Similarly to our derivation of the background equations of motion in section \ref{Background}, the more efficient technique we will employ is to treat, within the ADM formalism, the cosmological and the induced metric on equal footing.

Let us finally note that, although we will be ultimately interested in quasi de-Sitter backgrounds in the relativistic regime, our following analysis is exact and holds irrespective of the details of the homogenous dynamics.

\subsection{Second-order action for the perturbations}

We use the ADM formalism in which the metric is written in the form
\beq
g_{\mu \nu} d x^{\mu} dx^{\nu}=-N^2 dt^2 +h_{ij} (dx^i+N^i dt)(dx^j+N^j dt)\,
\label{g-ADM}
\eeq
where $h_{ij}$ is the spatial metric, $N^i$ is the shift vector and $N$ the lapse function. As discussed in the introduction of this section, following our bimetric perspective, it proves to be useful to introduce their induced gravity counterparts, such that
\beq
\tg_{\mu \nu} dx^{\mu} dx^{\nu}=-\tN^2 dt^2 +\th_{ij} (dx^i+\tN^i dt)(dx^j+\tN^j dt)\,.
\eeq
From Eqs. \refeq{induced-tg} and \refeq{g-ADM}, their explicit expressions are given by
\bea
\th_{ij}&=&h_{ij} +f G_{IJ} \partial_i \phi^I \partial_j \phi^J \label{h} \\
\tN_i& \equiv&  \th_{ij} \tN^j= h_{ij}N^j +f G_{IJ} \dot \phi^I \partial_i \phi^J \equiv N_i+ f G_{IJ} \dot \phi^I \partial_i \phi^J \label{Ni}  \\
\tN^2&=&  N^2 -f  G_{IJ} \dot \phi^I \dot \phi^J + \th^{ij} \tN_i \tN_j- h^{ij} N_i N_j \label{N}
\eea
where $\th^{ij}$ is the inverse of $\th_{ij}$ and $\phi^I$ is fully inhomogeneous. The ``gravitational'' part of the action (the first two terms in Eq. \refeq{action-f=cst}) can then be simply expressed as two copies of the Einstein-Hilbert action in the ADM form\footnote{When going 
from Eq. \refeq{action-f=cst} to Eq. \refeq{action-ADM-multifield}, we discarded some boundary terms, corresponding to the Gibbons-Hawking-York boundary action \cite{Gibbons:1976ue,York:1972sj} and its induced-gravity counterpart. The relevance of this kind of terms for Galileon-type interactions was recently discussed in Refs. \cite{Dyer:2008hb,Dyer:2009yg} in a slightly different context.}:
\bea
S_{{\rm grav}}&=&  \frac{\M^2}{2}  \int {\rm d}t \, {\rm d}^3 x \sqrt{h}   \left( N R^{(3)}+ \frac{1}{N}  (E_{ij} E^{ij}-E^2)  \right) \nn \\
&+& \frac{\tM^2}{2}  \int {\rm d}t \, {\rm d}^3 x \sqrt{\th}   \left( \tN \tR^{(3)}+ \frac{1}{\tN}  (\tE_{ij} \tE^{ij}-\tE^2)  \right)
\label{action-ADM-multifield}
\eea
where $\th=$ det$(\th_{ij})$, $\tR^{(3)}$ is the scalar Ricci curvature built from $\th_{ij}$, and the symmetric tensor $\tE_{ij}$ is defined by 
\beq
\tE_{ij}=\half \dot{\th}_{ij}-\frac12 \tilde{\nabla}_i \tN_j -\frac12 \tilde{\nabla}_j \tN_i  
\label{tEij}
\eeq
where $\tilde{\nabla}$ represents the covariant derivative associated to the metric $\th_{ij}$ (similar definitions hold of course for quantities without $\tilde{}$ ). It is then straightforward to derive the constraint equations for the lapse and the shift. Because the specific form of the brane action in \refeq{action-f=cst} is not important in this calculation, and because it actually simplifies the presentation, we derive these constraint equations for a general scalar field Lagrangian of the form $P(X^{IJ},\phi^K)$, where the kinetic terms $X^{IJ}$, defined in \refeq{def-XIJ}, take the ADM form
\be
X^{IJ}=\frac{1}{2 N^2}v^I v^J -\frac12 h^{ij} \partial_i \phi^I \partial_j \phi^J 
\ee
with
\be
v^I= \dot \phi^I-N^i \partial_i \phi^I \,.
\ee
The variation of the action with respect to $N$ then yields
the energy constraint
\beq
2(N^2 P-P_{\langle IJ \rangle} v^I v^J) + \M^2 \left( N^2 R^{(3)} +E^2-    E_{ij} E^{ij}  \right)+\tM^2 \sqrt{\frac{\th}{h}} \frac{N^3}{\tN^3}  \left( \tN^2 \tR^{(3)} +\tE^2-    \tE_{ij} \tE^{ij}  \right)
=0\, 
\label{N-constraint}
\eeq
while the variation of the action with respect to the shift $N_i$ gives 
the momentum constraint
\bea
&&\M^2 \nabla_j \left( \frac{1}{N} (E_i^j-E \delta_i^j) \right)-\frac{P_{\langle IJ \rangle}}{N} v^I \partial_i \phi^J +\tM^2 \sqrt{\frac{\th}{h}} \left[  h_{ik} \tilde{\nabla}_j \left( \frac{1}{\N} (\tE^{kj}-\tE \delta^{kj})  \right) 
\right.
 \cr
  && 
  \left.
+\frac12 \left( \tR^{(3)}-\frac{1}{\tN^2}(\tE_{ab} \tE^{ab}-\tE^2  ) \right) \frac{h_{ik}}{\tN}( \tN^k-N^k)
 \right]=0 \,
 \label{Ni-constraint}
\eea
where $P_{\langle IJ \rangle} \equiv \frac12 \left( \partial P/ \partial X^{IJ} +\partial P/ \partial X^{JI} \right) $. The remaining task is as follows: we decompose the lapse and shift as
\beq
N = 1 + \deltaN, \qquad  N_i = \beta_{,i}+V_i  \,
\eeq
where $\partial^i V_i=0$; we solve the linearized constraints equations for the auxiliary variables $\deltaN$, $\beta$ and $V_i$, insert their solutions back in the action and deduce the quadratic action in terms of the true propagating degrees of freedom: two tensor modes and $N$ scalar perturbations (there is no active source of vector modes). At linear order, these two types of fluctuations are decoupled and can be treated separately. We consider scalar perturbations in the next two subsections and then move on to discuss gravitational waves. In practice, the intermediate calculations for the scalar sector can be simplified by making appropriate choices of gauge. In this respect, before discussing the multifield situation, it is instructive to consider the case of a single inflaton field ($N=1, G_{11}=1$), to which we now turn.

\subsubsection{S\lowercase{ingle field case}}

A particularly convenient gauge in the single field case is the uniform inflaton gauge in which the (single) scalar perturbation $\z$ appears in the spatial metric $h_{ij}$ in the form $h_{ij}=a(t)^2 e^{2 \z} \delta_{ij}$ while the inflaton is homogeneous $\phi=\phi(t)$. The induced quantities \refeq{h}, \refeq{Ni}, \refeq{N} then considerably simplify to $\th_{ij}=h_{ij}$, $\tN_i=N_i$ and $\tN^2=N^2-1+\c^2$. Details of the calculations for a general $k$-inflationary Lagrangian of the form $P(X,\phi)$, with $X=-\frac12 g^{\mu \nu} \partial_{\mu} \phi \partial_{\nu} \phi$, can be found in appendix \refeq{Linear}. The resulting second-order (scalar) action is of the form 
\bea
S_{(2)}= \int {\rm d}t\,    {\rm d}^3 x \,a^3  \left(  A(t) {\dot \zeta}^2 -B(t) \frac{(\partial \zeta)^2}{a^2}   \right)\,,
\label{2d-order-action-2}
\eea
which simply expresses the fact that $\zeta$ is exactly massless and hence conserved on large scales -- in case the decaying mode is indeed decaying -- (see \cite{Naruko:2011zk,Gao:2011mz} for a non-linear proof in a broader context). Concentrating on the DBI Galileon case, $A$ and $B$ can be cast in the following simple form:
\bea
A(t)&=& \frac{\M^2}{\c^2} \left( \eps \, \X^2+3\c^2(1-\X^2) \right) + \frac{\tM^2}{\c^3} \left( \teps \, \X^2+3\c^2\left(1-\frac{\X^2}{\c^4}\right) \right) \,,\label{A}  \\
B(t) &=& \M^2 \left( \eps \, \X (3 \, \X-2)+\X-1  \right) +\frac{\tM^2}{\c} \left( \teps \, \X \left(3 \,\frac{ \X}{\c^2}-2\right)+\X-\c^2 \right)\,,
\label{B}
\eea
where we remind the reader that $\teps$, defined in Eq. \refeq{teps}, is the deceleration parameter associated to the background induced metric and $\X$ was defined in Eq. \refeq{def-kappa}. Hence, this formulation reveals how the fluctuation $\zeta$ reacts to the two background geometries: without the induced gravity, $\tM^2=0$, $\X=1$ and one recovers the standard $k$-inflationary result \cite{Garriga:1999vw} with a speed of sound $B(t)/A(t)=\c^2$. At the other extreme limit, if one formally considers $\M^2=0$, then $\X=\c^2$ and one finds a similar expression
\bea
S_{(2),\M^2=0}= \int {\rm d} \tt\, {\rm d}^3 x \,a^3 \, \teps \, \tM^2 \c^2  \left(   \left(\frac{d \zeta}{d \tt}\right)^2 -\frac{1}{\c^2} \frac{(\partial \zeta)^2}{a^2}   \right) 
\label{S2-MP=0}
\eea
in terms of the quantities associated to the induced metric with the replacement 
$\c^2 \to \c^{-2}$. This can be easily understood: one can then treat the metric $\tg$ as the cosmological metric\footnote{This argument is not completely straightforward and can not be applied blindly to the multifield case for example. Details about its validity can be found in appendix \refeq{Induced gravity}.}, in which case the action for gravity becomes canonical while the brane action can be expressed as $S_\matter =  \int d^4 x \sqrt{-\tg} {\tilde P}(X_{\tg},\phi)$ where $X_{\tg} \equiv -\frac{1}{2}\tg^{\mu \nu}\partial_{\mu} \phi \partial_{\nu} \phi $ and
\bea
{\tilde P}= \left( \frac{1}{f}-V \right) \sqrt{1+2 f X_{\tg}}  -\frac{1}{f}\,.
\label{Pmodified}
\eea
Using the general expression derived in \cite{Garriga:1999vw}, it is easy to verify that the speed of sound is given by $1/\c^2$  for the Lagrangian \refeq{Pmodified}, with the consequence that the speed of sound with respect to the cosmological cosmic time is unity. In the general situation, one can not define an Einstein frame in which the gravitational action becomes canonical and one must resort to the full calculation, leading to the result \refeq{2d-order-action-2} that nicely interpolates between the two extreme cases aforementioned, the relative importance of the two geometries being determined by the parameter $\X$. However, the form \refeq{A} of the kinetic term is not appropriate for discussing the possible presence of a ghost. Because the conditions for avoiding the presence of ghosts are, a priori, more restrictive in the multifield case, we now turn to this situation, restricting our attention to two fields ($I=1,2$) and a trivial field space metric $G_{IJ}=\delta_{IJ}$ for simplicity of presentation.

\subsubsection{M\lowercase{ultifield case}}

Contrary to the previous subsection, the fluctuations of all the scalar fields can not be erased in the multifield case by a suitable choice of time-slicing. Instead, a convenient gauge is the spatially flat gauge, in which the spatial metric takes its unperturbed value -- $h_{ij}=a^2(t) \delta_{ij}$ -- while the scalar degrees of freedom are directly taken to be the scalar fields fluctuations $Q^I$ such that $\phi^I={\bar \phi}^I(t)+Q^I(t,\boldsymbol{x})$ (we omit the bar on ${\bar \phi}^I$ in the following when there is no ambiguity). Details of the linearized constraint equations and their solutions can be found in appendix \refeq{Linear}. Intermediate steps in the calculation of the second-order action are very lengthy and not illuminating. However, the final exact result can be cast in a very useful and elegant form. For that purpose, it is convenient to decompose the scalar field fluctuations into  
\be
Q^I=Q_\s e^I_\s+Q_s e^I_s\,,
\ee
where $e^I_{\s} \equiv \dot \phi^I/\sd$ is the unit vector (with respect to the field space metric) pointing along the background trajectory in field space and $e_s^I$ is the unit vector orthogonal to $e^I_\s$. In this way, the so-called adiabatic perturbation $Q_{\s}$ inherits all the properties of the singe field case while $Q_s$, called the (instantaneous) entropy perturbation, embodies the genuinely multifield effects \cite{Gordon:2000hv} (there would be $N-1$ entropy perturbations in general). Indeed, entropy fluctuations are fluctuations away from the background trajectory, something that can not happen in the single field case in which fluctuations only represent time-delays along the background trajectory.

After going to conformal time $\tau = \int {{\rm d}t}/{a(t)}$, and, upon using the canonically normalized fields $v_{\s} \equiv   z\frac{H}{\sd} \, Q_{\s}$ and $v_{s} \equiv  \w \, Q_s$ with 
\bea
&&z= \frac{a}{\c^{3/2}} \frac{\sd}{H} \left((1-9 \q)\X^2+6 \q \X \c^2  \right)^{1/2} \label{z} \,,\\
&&\w=\frac{a}{\sqrt{\c}}\,\left(1-3\q\right)^{1/2}  \label{w}\,,
\eea
the second-order action can be expressed in the remarkably simple form:
\begin{eqnarray}
\label{S_v}
S_{(2)}&=&\frac{1}{2}\int {\rm d}\tau\,  {\rm d}^3x \Big\{ 
  v_\s^{\prime\, 2}+ v_s^{\prime\, 2} -2\xi v_\s^\prime v_s-\ca^2  (\partial v_\s)^2 -\ce^2(\partial v_s)^2  -2 \cae \partial v_\s \partial v_s
\cr
&& 
\left.
\qquad
+\frac{z''}{z} v_\s^2
+\left(\frac{\w''}{\w}-a^2 \mu_s^2\right) v_s^2+2\, \frac{z'}{z}\xi v_\s v_s\right\}
\end{eqnarray}
where a prime denotes a derivative with respect to conformal time and 
\begin{eqnarray}
\label{11}
&&\ca^2 =  \left( (1-9 \q)\X+6 \q \c^2  \right)^{-1}   \left[ \frac{\c^2}{\X} \left( (1-5 \q)\X+2 \q \c^2 \right)  +6 \q s \c^2 \frac{\X-\c^2}{1-\c^2}    \right] \label{ca} \,,
\\
&&\ce^2= \frac{\c^2}{\X} \left[1+ \frac{2 s \q}{1-3 \q}  \frac{\X-\c^2}{1-\c^2}    \right] \label{ce} \,,
\label{z,a} \\
&&\cae=\frac{  \c^3 (1-\X) (1-\c^2)}{\left(  \X (1-3\q)^3  \left( (1-9\q) \X+6\q \c^2 \right)\right)^{1/2}} \frac{2 \M^2 HV_{,s}}{\sd^3} \,, \\
&&\mu_s^2= \frac{ \c V_{,ss}}{(1-3 \q)}    - \frac{\left((1 - 9 \q) \X^2 + 2 (\X-1+3 \q ) \c^2\right)}{(1-3 \q)^3} \frac{V_{,s}^2}{\sd^2} \,, \\
&&\xi=-a  \frac{\X \left((1-9\q)\X+\c^2(1+3 \q) \right)}{  \left( (1-3\q)^3  \left( (1-9\q) \X^2+6\q \X \c^2 \right)\right)^{1/2} } \frac{V_{,s}}{\sd}\,\label{def-xi}
\eea
where $V_{,s} \equiv e_s^I V_{,I}$ and $V_{,ss}\equiv e_s^I e_s^J V_{,IJ}$ are gradients of the potential projected along the entropy direction in field space (and similarly in the following). The resulting equations of motion read, in Fourier space:
\begin{eqnarray}
\hspace*{-1.5em}
 &&v_{\s}''-\xi v_{s}'+\left(\ca^2 k^2-\frac{z''}{z}\right) v_{\s} -\frac{(z \xi)'}{z}v_{s}+\cae k^2 v_s=0\,,   \qquad
\label{eq_v_sigma}
\\
\hspace*{-1.5em} 
 &&v_{s}''+\xi  v_{\s}'+\left(\ce^2 k^2- \frac{\w''}{\w}+a^2\mu_s^2\right) v_{s} - \frac{z'}{z} \xi v_{\s}+\cae k^2 v_{\sigma}=0  \,
\label{eq_v_s}
\end{eqnarray}
where one can check that standard multifield DBI inflation \cite{Langlois:2008qf} is recovered in the limit $\q=0,\X=1$. Similarly to the single-field case, one can also understand in a simple fashion the other extreme limit, in which one formally considers $\M=0$, by considering $\tg$ as the cosmological metric. The proof, more subtle than in the single field-case, can be found in appendix \refeq{Induced gravity}.

Let us now analyze the physical consequences of the above result. First, by taking the square roots in \refeq{z}-\refeq{w}, we implicitly consider the regime in which the fluctuations are not ghosts, which reads 
\be
\q < (9-6 \c^2/\X)^{-1}\,.
\label{alpha-bound}
\ee
In particular, $\q <1/3$ in the ghost-free regime, as we said in our discussion of the background evolution in section \refeq{Background}. Interestingly, using its definition \refeq{def-alpha}, the condition \refeq{alpha-bound} on $\q$ can be reformulated, through the first Friedmann equation \refeq{Friedmann1}, as an upper bound on the energy density $\rho_\matter$, \textit{i.e.} there exists a critical background energy density 
\be
\rc =  \left(1+\frac{\c^3 \M^2}{\tM^2}\right) /    \left( f \c \left(3-2\frac{\c^2}{\X}\right) \right)
\label{rho-c}
\ee
above which the cosmological fluctuations become ghosts. This novel aspect due to the induced gravity can be simply understood in the formal limit $\M \to 0$. Indeed, the bound \refeq{rho-c} then boils down to $V < 1/f$, which is a simple consequence of the form of the kinetic term in \refeq{Pmodified}. 

Another point worth stressing is the very simple form of the coupling between adiabatic and entropy perturbations on super-horizon scales, \textit{i.e.} when spatial gradients can be neglected: it is given by a single parameter $\xi$ \refeq{def-xi}. Note also that this property is not particular to DBI Galileon inflation but is shared by a very large class of multifield inflationary models \cite{Langlois:2008mn}. One novel feature however, due to the common presence of both the Einstein-Hilbert and induced gravity action, is the gradient coupling between adiabatic and entropy perturbations through the term $\cae$ in \refeq{eq_v_sigma}-\refeq{eq_v_s}. When it can not be neglected, it implies that one must diagonalize the speed of sound matrix and that the adiabatic and entropy perturbations can not be quantized independently under the horizon (nor the field fluctuations themselves). In the following, we consider a regime in which this coupling can be neglected, in which case $\ca^2$ and $\ce^2$ are truly the respective speed of propagation squared of adiabatic and entropy fluctuations. Because $s \equiv \dot{c}_\D/(H \c)$, which can have a priori any sign and any value, enters into their explicit expression \refeq{ca} and \refeq{ce}, $\ca^2$ and $\ce^2$ have no definite signs in general. With a slow-varying approximation in mind, we neglect $s$ in the following, in which case one can check that $\ca^2$ and $\ce^2$ are both positive definite in the ghost-free regime, \textit{i.e.} there is no gradient instability. In this regime, the entropic speed of sound $\ce$ is always less than the adiabatic one $\ca$, the only particular cases when they become equal being:
\begin{itemize}
\item standard multifield DBI inflation ($\tM \to 0$ and hence $\q \to 0$ and $\X \to 1$). One then have $\ca=\ce=\c$ \cite{Langlois:2008wt}.
\item the non-relativistic limit of DBI Galileon ($\c \to 1$ for any $\tM$). One then have $\ca=\ce  =1$.
\item the limit in which the induced gravity dominates ($\M^2 \ll \tM^2/\c$ and hence $\X \simeq \c^2$). One then have $\ca=\ce \simeq 1$ for any $\c$.
\end{itemize}
Note also that $\ce^2 \simeq \c^2/\X$ is always less than one, contrary to the adiabatic speed of sound. If we require the latter to be subluminal, one should impose the condition $\q \leq  (9-2\c^2/\X)^{-1}$, which is more stringent than the condition \refeq{alpha-bound} for avoiding ghosts. Finally, by comparing the exact second-order action \refeq{S_v} with the one obtained from the procedure consisting in only perturbing the scalar fields, and not the metric:
\bea
S_{(2),{\rm naive}}&=& \half \int {\rm d}t \,\dn{3}{x}\,    a^3 \left( \frac{\dot Q_{\sigma}^2}{\c^3}\left(1-3\q (3-2\c^2) \right)-\frac{(\partial Q_{\sigma})^2}{\c a^2}\left( 1- \q (5-2\c^2) \right)
\right.
\cr
&&  \hspace*{+6.5em} 
\left.
+\frac{1-3\q}{\c} \left( \kineticperp-\c^2 \frac{\gradperp}{a^2}  \right)+\ldots
  \right)\,,
    \label{S2flat}
\eea
we see that the latter is too naive and leads to completely wrong results (for example for the speeds of propagation) as soon as $\X$ differs significantly from one. This is in agreement with the structure of the solutions to the constraints equations in the flat gauge in appendix \refeq{Linear}.

\subsubsection{Gravitational waves}

Here, we derive the second-order action for the tensor modes $\E_{ij}$, \textit{i.e.} the transverse -- $\partial^i \E_{ij}=0$ -- and trace free -- $\E^i_i=0$ -- part of $h_{ij}/a^2$, where $h_{ij}$ is the spatial metric in \refeq{g-ADM}. This is actually straightforward within the bimetric approach. We know the result corresponding to the standard Einstein-Hilbert action:
\beq
S_{(2), EH} = \sum_\po \int {\rm d} t \,{\rm d}^3 \bk \frac{a^3  \M^2}{8} \left[ \left( \dot \E_{\bk}^\po\right)^2 - \frac{k^2}{a^2}\left( \E_{\bk}^\po \right)^2\right]\, 
\eeq
where
\beq
 \E_{ij} = \int \frac{{\rm d}^3 \bk}{(2\pi)^3} \sum_{\po=+,\times} \epsilon^\po_{ij}(k) \E^\po_{\bk}(\tau) e^{i {\bk} \cdot {{\boldsymbol x}}}\
\eeq
is the standard decomposition of $\E_{ij}$ into $+$ and $\times$ polarization states. Now, as far as the tensor sector is concerned, the Einstein-Hilbert and induced gravity and action are equivalent, with the replacements $\M \leftrightarrow \tM$ and $dt  \leftrightarrow d\tt $, where $\tt$ is the cosmic time associated to the induced metric, such that $d \tt =\c d t$ \refeq{induced-time}. Hence, one immediately deduces from this the total second-order action for the tensor modes
\beq
S_{(2),t} =  \sum_\po \int {\rm d} t \,{\rm d}^3 \bk \frac{a^3}{8} \left[ \left(\M^2+\frac{\tM^2}{\c}\right) \left( \dot \E_{\bk}^\po\right)^2 - \frac{k^2}{a^2} \left( \M^2+\tM^2 \c\right)  \left( \E_{\bk}^\po \right)^2\right]\, 
\eeq
(this result actually also holds for a general, non constant, warp factor). In terms of the canonically normalized fields in conformal time
\beq
v_{\bk}^\po \equiv \frac{a}{2} \sqrt{\M^2+\frac{\tM^2}{\c}}  h_{\bk}^\po\, ,
\label{SM-tensor}
\eeq
this reads
\beq
\label{equ:S2v3}
S_{(2),t} = \sum_\po \frac{1}{2} \int {\rm d} \tau {\rm d}^3 \bk  \left[ (v^\po_{\bk}{}')^2   - \left(\cg^2  k^2 - \frac{\qg''}{\qg} \right) (v_{\bk}^\po)^2\right] \, ,
\eeq
where $q=a \sqrt{\M^2+\tM^2/\c}$ and 
\be
\cg^2 \equiv \frac{ \M^2+\tM^2 \c}{\M^2+\tM^2/\c}
\ee
is the speed of propagation squared of the gravitational waves, which is always subluminal.

\subsection{Quantization and power spectra}

The exact calculation of the second-order action above made it possible to identify the remarkably simple form of the linear equations of motion \refeq{eq_v_sigma}-\refeq{eq_v_s}. In the remaining of this paper, we restrict ourselves to considering a quasi de-Sitter inflationary phase -- $a(\tau) \simeq -\frac{1}{H \tau}$ -- in a slow-varying regime in which the time evolution of every quantity is slow with respect to that of the scale factor, such that for example $z''/z\simeq \w''/\w \simeq q''/q \simeq 2/\tau^2$. As explained above, we neglect the gradient coupling between adiabatic and entropy perturbations, and we assume that the effect of the coupling $\xi$ can be neglected on sub-horizon scales, so that adiabatic and entropy perturbations can then be quantized independently. We also consider an effectively light entropy perturbation, such that $|\mu_s^2|/H^2\ll 1$, and which therefore gets amplified at (its) sound-horizon crossing\footnote{Note that multifield models of inflation with intermediate masses of order $H$ can have a rich phenomenology \cite{Chen:2009zp,Chen:2009we}, especially at the non-linear level.}.

The solutions to Eqs. \refeq{eq_v_sigma} and \refeq{eq_v_s} corresponding to the usual vacuum on very small scales then read
\bea
v_{\s\, k} & \simeq &   \frac{1}{\sqrt{2k \ca}}e^{-ik \ca \tau }\left(1-\frac{i}{k \ca \tau}\right)\,, \\
v_{s\, k}  & \simeq &   \frac{1}{\sqrt{2k \ce}}e^{-ik \ce \tau }\left(1-\frac{i}{k \ce \tau}\right) \,,
\eea
form which one deduces, using \refeq{z}-\refeq{w} and \refeq{ca}-\refeq{ce}, the power spectra of the initial adiabatic and entropy fields:
\bea
{\cal P}_{\Qsi^*} & \simeq &\left( \frac{H}{2 \pi} \right)^2   \frac{\left((1-9\q)\X^2+6 \q \X \c^2 \right)^{1/2}}{\left((1-5\q)\X+2\q \c^2\right)^{3/2}} \label{PQsi} \\
{\cal P}_{\Qs^*} &\simeq &  \left( \frac{H}{2 \pi \c} \right)^2 \frac{\X^{3/2}}{1-3\q}\,. \label{PQs}
\eea
Strictly speaking, the power spectra in Eqs. \refeq{PQsi}-\refeq{PQs} are evaluated respectively soon after adiabatic and entropic sound-horizon crossing. Here, we have assumed that the hierarchy between the two sound speeds is not too large, so that the parameters entering into Eqs. \refeq{PQsi}-\refeq{PQsi}, such as $H$ and $\q$, do not vary too much between the two epochs. Note that this is not as restrictive as it may seem, as the number of efolds between them depends only logarithmically on the ratio $\ca/\ce$. From the above results, one deduces that the enhancement of the entropy fluctuations with respect to the adiabatic ones that was found in standard multifield DBI inflation \cite{Langlois:2008wt} -- for which $Q_s \sim Q_{\sigma}/\c$ -- is even more important in multifield DBI Galileon inflation, namely ${\cal P}_{\Qs^*} \geq \frac{1}{\c} {\cal P}_{\Qsi^*}$. Finally, rewriting the above result \refeq{PQsi} in terms of the curvature perturbation
 \be
{\R}=-\frac{H}{\dot \s}Q_{\s}\,,
\label{zeta}
\ee
one finds its power-spectrum around horizon crossing
\be
{\cal P}_{\R_*}= \left( \frac{H}{\sd} \right)^2  \left( \frac{H}{2 \pi} \right)^2   \frac{\left((1-9\q)\X^2+6 \q \X \c^2 \right)^{1/2}}{\left((1-5\q)\X+2\q \c^2\right)^{3/2}}  \,.
\label{power-spectrum-R}
\ee
In the relativistic regime of particular interest to us, we have seen in section \refeq{Background} that a phase of quasi de-Sitter expansion is achievable under the conditions \refeq{condition1} and \refeq{condition2}. In this regime, the power spectrum \refeq{power-spectrum-R} takes the form
\be
{\cal P}_{\R_* \,\, \c \ll 1}=  \frac{1}{8 \pi^2 \ca \eps}\left( \frac{H}{\M}\right)^2  \frac{  (1-3\q)}{\left(1-5\q\right)}  \,.
\label{power-spectrum-R}
\ee

The determination of the tensor power spectrum runs analogously. In the slow-varying regime, the Bunch-Davies vacuum solution of Eq. \refeq{equ:S2v3} reads
\bea
v_{\bk}^\po & \simeq &   \frac{1}{\sqrt{2k \cg}}e^{-ik \cg \tau }\left(1-\frac{i}{k \cg \tau}\right)\,, 
\eea
from which one deduces, together with Eq. \refeq{SM-tensor}:
\be
{\cal P}_{t} = \frac{2H^2}{\pi^2} \frac{(\M^2+\tM^2/\c)^{1/2}}{(\M^2+\tM^2 \c)^{3/2}}\,.
\ee
In the relativistic regime, in which the condition $\tM^2 \ll \M^2 \c^3$ should hold to achieve an inflationary phase, the effect of the induced gravity is therefore subdominant, and the power-spectrum of the gravitational waves is not modified at leading order compared to its result in Einstein gravity.

\subsection{Large scale evolution}

In the single field case, relating the quantum fluctuations generated during inflation to the primordial seeds of the large-scale structure deep in the radiation era relies on the constancy of the curvature perturbation $\z$ on super horizon scales. As is well known, and shown for the first time in \cite{Starobinsky:1994mh}, this property does not hold in general in a multifield situation due to the large scale feeding of the curvature perturbation by the entropic fluctuations. This can easily be seen here by noticing the existence of an exact first-order integral of the adiabatic equation \refeq{eq_v_sigma} when spatial gradients can be neglected\footnote{This result can also be obtained from the large-scale limit of the energy constraint \refeq{energy} together with the momentum constraint \refeq{momentum}.}:
\be
\left( \frac{v_\s}{z} \right)' \simeq \frac{\xi}{z} v_s \Leftrightarrow \zeta' \simeq -\frac{\xi}{z}v_s\,.
\ee
As for the entropic perturbation, one can use this result, together with Eq. \refeq{eq_v_s}, to show that it evolves independently of the curvature perturbation on large scales:
\be
v_{s}''+\left(- \frac{\w''}{\w}+a^2\mu_s^2  +\xi^2 \right) v_{s} \simeq 0\,.
\ee
Quite generally, upon neglecting the decaying modes, one can therefore express the late-time curvature perturbation as
\be
\R \simeq \A_\sigma  Q_{\s*} + \A_s  Q_{s*} 
\label{R-fields}
\ee
where
\be
\A_\sigma = -\left( \frac{H}{\dot \s}\right)_*  \qquad
\A_s =  \T \left( \frac{ \c H}{\dot \s}  \frac{(1-3\q)^{1/2} ((1-9\q)\X^2+6\q \X \c^2)^{1/4}}{ ((1-5\q)\X^2+2\q \X \c^2)^{3/4}}  \right)_*  
\label{comparision-transfer}
\ee
and the complicated expression of $\A_s$ has been chosen such that
\be
{\cal P}_{\R}= {\cal P}_{\R_*}   \left( 1+\T^2 \right)\,,
\label{power-spectrum-R-final}
\ee
\textit{i.e.} $\T^2$ measures the contribution of the entropy perturbations to the final curvature perturbation. In the following, we concentrate on the curvature perturbation and leaves aside the interesting possibility of a remaining primordial isocurvature perturbation.

\section{N\lowercase{on-Gaussianities}}
\label{NGs}

This section is devoted to the analysis of the primordial bispectrum generated by multifield DBI Galileon inflation in the relativistic regime. Schematically, primordial non-Gaussianities can be generated at three different epochs of the inflationary evolution:
\begin{itemize}
\item before the relevant modes cross their respective sound horizon. The fluctuations are then in their vacuum and their rapid oscillations usually average out. However, in cases in which the background evolution displays sharp or periodic features, constructive interferences between different modes can be generated, and hence significant non-Gaussianities \cite{Chen:2006xjb,Chen:2008wn,Leblond:2010yq,Flauger:2010ja,Chen:2010bk}. 
\item around horizon crossing. This contribution is unavoidable and is typically large in models in which derivative operators are important, like in scenarios with non-standard kinetic terms (see e.g the review \cite{Chen:2010xk}).
\item after horizon crossing. The curvature perturbation being non-perturbatively constant on large scales in single-field models \cite{Lyth:2003im,Rigopoulos:2003ak,Lyth:2004gb,Langlois:2005ii,Langlois:2005qp,Naruko:2011zk,Gao:2011mz}, this mechanism is only possible in multifield scenarios. The resulting shape of the non-Gaussianities depends on the masses of the entropy fluctuations while its amplitude is very model-dependent (see e.g \cite{Elliston:2011dr} for a recent illustration).
\end{itemize}
Following our analysis of the linear fluctuations, we consider a smooth quasi de-Sitter background evolution and hence do not consider the first type of non-Gaussian contributions. As for the ones after horizon crossing, they would be of the well known local type as we consider en effectively light entropy perturbation. Their amplitude being model-dependent, we do not consider them in the following and concentrate on the second type of contributions (Ref. \cite{RenauxPetel:2009sj} gives an example, in the DBI context, in which local-type non-Gaussianities and ones generated around horizon-crossing can both be significant).

\subsection{Third-order action}
\label{3rd}

To determine the bispectrum generated by the quantum interferences amongst different modes of perturbations, we should calculate the third-order action in these perturbations. Fortunately, in the almost de-Sitter and relativistic regime of interest (see section \ref{Background}), we do not need its exact form. Indeed, using the results of appendix \refeq{Linear}, one can check that, in the flat gauge, the metric perturbations are negligible compared to the field fluctuations themselves. Therefore, it is sufficient for our purpose to perturb the scalar fields only and at leading order in the slow-varying approximation. Hence, one can consider separately the brane action \refeq{brane-action} and the induced gravity action. The corresponding result for the former, determined in \cite{Langlois:2008qf}, is given by
\bea
S_{(3)}^{\rm DBI}&=& \int {\rm d}t \,\dn{3}{x}\, a^3    \frac{ f \sd}{2 \c^5 }  \left( \dQsi^3 - \c^2 \dQsi \frac{(\pQsi)^2}{a^2}
\right.
\cr
&&
\left.
+\c^2 \dQsi \kineticperp+\c^4 \dQsi  \frac{\gradperp}{a^2} -2 \frac{\c^4}{a^2} \pQsi \kineticgradperp
\right)\,.
\eea
For the induced gravity action, it proves very useful, like for the second-order action, to start from its ADM form Eq. \refeq{action-ADM-multifield}. Here, we simply give the result, leaving details of the calculation in appendix \refeq{Perturbations}:
\bea
S_{(3)}^{\rm ind}&=& \int {\rm d}t \,\dn{3}{x}\,    a^3 f^2 H^2 \tM^2 \sd \left(  -\frac{3}{\c^7}(5-2\c^2) \dQsi^3 +\frac{9-2\c^2}{\c^5}\dQsi \frac{(\pQsi)^2}{a^2}
\right.
\cr
&&
\left.
+\frac{6}{H \c^5} \dQsi^2 \frac{ \LapQsi }{a^2}-\frac{3}{2 H \c^3}\frac{(\pQsi)^2}{a^2} \frac{\LapQsi}{a^2}
\right.
\cr
&&
\left.
-\frac{9}{\c^5}\kineticperp \dQsi
+\frac{1}{\c^3}\frac{\gradperp}{a^2} \dQsi
+\frac{6}{a^2 \c^3} \pQsi \kineticgradperp
+\frac{2}{H\c^3} \kineticperp \frac{\LapQsi}{a^2}
\right.
\cr
&&
\left.
+\frac{4}{H \c^3} \dQsi \frac{\doneLaponeperp}{a^2}
-\frac{1}{2H\c} \frac{\gradperp}{a^2} \frac{\LapQsi}{a^2}
-\frac{1}{a^4 H \c} \pQsi \pQs \LapQs
 \right)
 \label{S3ind}
\eea
For completeness, we have kept the terms subleading in $\c^2$ that appear in the calculation (in the first line). However, we should bear in mind that terms not calculated here, coming from the mixing with gravity or beyond leading order in the slow-varying approximation, can be of the same order of magnitude or greater as those terms (see for instance \cite{Chen:2006nt,Burrage:2011hd}).

Note that, at leading order in a slow varying approximation, only shift-symmetric operators can appear, and only with an even number of entropic fields $Q_s$. Note also that the seven mixed adiabatic/entropic operators are the multifield generalisations of the four purely adiabatic ones. Eventually, one can check that the relation between the adiabatic speed of sound $\ca$ and the coefficient of the operator $\dQsi (\pQsi)^2$ that was noted in the context of the effective field of theory of inflation \cite{Cheung:2007st} is verified. However, because of the presence of higher-order vertices, the direct link between $\ca$ and the amplitude of the equilateral-type non-Gaussianities is spoiled, as has already been noted in \cite{Burrage:2010cu}. To see this, let us use the (leading-order) equations of motion 
\bea
&&\ddot{Q}_{\sigma}  +3H \dQsi-\ca^2 \frac{\partial^2 \Qsi}{a^2}=0 \\
&&\ddot{Q}_{s} +3H \dQs-\ce^2 \frac{\partial^2 \Qs}{a^2}=0\,
\eea
to simplify the third-order action\footnote{The fact that this procedure enables one to calculate accurately correlation functions (if boundary terms are carefully taken into account) was discussed in Refs. \cite{Seery:2005gb,Seery:2006tq,Seery:2010kh}, more recently in Refs. \cite{Arroja:2011yj,Burrage:2011hd,Rigopoulos:2011eq} and applied to the most general scalar-tensor theory with second order equations of motion in \cite{RenauxPetel:2011sb}.}. Using integrations by part, one finds that
\bea
\int {\rm d}t \,\dn{3}{x}\, a^3 \, \dQsi^2 \frac{\LapQsi}{a^2}&=&  \int {\rm d}t \,\dn{3}{x}\, a^3  \left( \frac{2H}{\ca^2} \dQsi^3\right) \label{1} \\
\int {\rm d}t \,\dn{3}{x}\, a^3 \,\frac{1}{H} \frac{ \LapQsi}{a^2} \frac{(\pQsi)^2}{a^2}&=&  \int {\rm d}t \,\dn{3}{x}\, a^3  \left( \frac{2}{\ca^4} \dQsi^3+\frac{2}{\ca^2} \dQsi \frac{(\pQsi)^2}{a^2}\right) \label{2} \,.
\eea
Hence, one can trade the two higher dimension operators $\dQsi^2 \LapQsi$ and $(\pQsi)^2 \LapQsi$ in the second line of \refeq{S3ind} for the standard $k$-inflationary operators $\dQsi^3$ and $\dQsi (\pQsi)^2$, as was noted in Ref. \cite{Creminelli:2010qf}. Similarly, for the mixed adiabatic/entropic operators, one can show that
\bea
\int {\rm d}t \,\dn{3}{x}\, a^3 \, \dQsi \dQs  \frac{\LapQs}{a^2}&=&  \int {\rm d}t \,\dn{3}{x}\, a^3  \left( \frac{3H}{\ce^2} \dQsi \dQs^2-\frac{1}{2 \l^2} \dQs^2 \frac{\LapQsi}{a^2}\right)  \label{3}
\eea
and
\bea
\int {\rm d}t \,\dn{3}{x}\, a^3 \,\frac{1}{H} \frac{ \pQsi \pQs}{a^2} \frac{\LapQs}{a^2}&=&  \int {\rm d}t \,\dn{3}{x}\, a^3 \frac{1}{\ce^2} \left[ 
 \dQsi  \frac{\gradperp}{a^2} 
 + \frac{2}{a^2} \pQsi \kineticgradperp
 \right.
\cr
&& \hspace*{-5.0em}
\left.
+\frac{1}{2H} \kineticperp \frac{\LapQsi}{a^2}
+\frac{1}{H} \dQsi \frac{\doneLaponeperp}{a^2}
-\frac{\ca^2}{2H} \frac{\gradperp}{a^2} \frac{\LapQsi}{a^2}
\right]  \label{4} \,
\eea
where we have introduced the ratio between the entropic and the adiabatic speed of sound
\be
\l \equiv  \frac{\ce}{\ca} \simeq \sqrt{ \frac{1-3\q(3-2\c^2)}{1- \q(5-2\c^2)}  }\,.
\label{def-beta}
\ee
Using the four equalities \refeq{1}-\refeq{4}, one can then write down the effective total third-order action as
\bea
S_{(3)}^{\rm eff}&=& \int {\rm d}t \,\dn{3}{x}\,  a^3    \frac{ f \sd}{2 \c^5 } \left[ A_{\dQsi^3}   \dQsi^3 
 - \c^2 A_{\dQsi (\pQsi)^2}  \dQsi \frac{(\pQsi)^2}{a^2}
\right.
\cr
&&
\left.
+\c^2 \dQsi \kineticperp+\c^4 \dQsi \frac{(\pQs)^2}{a^2}-2 \c^4 (1-2\q) \frac{\pQsi \pQs}{a^2} \dQs  
\right.
\cr
&&
\left.
-\frac{3}{2} \q \c^4\left( \frac{1}{\l^2}-1 \right)  \kineticperp \frac{\LapQsi}{H a^2}
+\frac{ \q}{2} \c^6\left( \frac{1}{\l^2}-1 \right) \frac{\gradperp}{a^2}  \frac{\LapQsi}{H a^2}
\right]
\label{S3final}
\eea
with
\bea
A_{\dQsi^3} & =&1-3\q \left(5-2\c^2-4 \l^2+\l^4 \right) \label{Afirst} \\
A_{\dQsi (\pQsi)^2} &=& 1-\q \left( 9-2 \c^2-3 \l^2 \right)\,. \label{Asecond}
\eea
As we will see, the correlation functions induced by each of the seven operators in \refeq{S3final} are linearly independent. Hence, no integration by part can simplify this result further, \textit{i.e.} none of the operators is redundant \cite{RenauxPetel:2011sb}. In particular, contrary to the purely adiabatic perturbations, one can not get rid of all the higher-order operators in the multifield scenario, hence the last line in Eq. \refeq{S3final}. 

At lowest order and at tree level, one deduces the bispectrum from the third-order action as \cite{Weinberg:2005vy}
\bea
\langle \R (t,\boldsymbol{k}_1)  \R (t,\boldsymbol{k}_2)  \R (t,\boldsymbol{k}_3)\rangle =-i \int_{-\infty}^t {\rm d}t' \langle 0 | \R^I (t,\boldsymbol{k}_1)  \R^I (t,\boldsymbol{k}_2)  \R^I (t,\boldsymbol{k}_3) H_I(t')  | 0 \rangle + {\rm c.c}
\eea 
where $\z^I$ is in the interaction picture, $H_I$ denotes the interaction Hamiltonian, such that $S_{(3)}^{\rm eff}=-\int {\rm d}t' H_I(t')$, and $  | 0 \rangle $ is the free theory vacuum. In practice, one evaluates this integral using the conformal time $\tau \simeq -1/(aH)$, in which case one can extrapolate the upper bound to $\tau= 0$ as the main contribution to the integrand comes from the period around horizon crossing. One also uses the appropriate rotated contour in the complex plane $(\tau \to -(\infty-i \eps))$. Following this procedure, and using Eq. \refeq{R-fields} and the structure of $S_{(3)}^{\rm eff}$, the bispectrum is calculated as
\begin{eqnarray}
\langle \R (\boldsymbol{k}_1)  \R (\boldsymbol{k}_2)  \R (\boldsymbol{k}_3)\rangle &=& (\A_{\s})^3 \langle Q_{\s}  (\boldsymbol{k}_1) Q_{\s}  (\boldsymbol{k}_2)  Q_{\s}  (\boldsymbol{k}_3) \rangle \nn \\
&+&\A_{\s}(\A_{s})^2(\langle Q_{\s}  (\boldsymbol{k}_1)  Q_s (\boldsymbol{k}_2)  Q_s (\boldsymbol{k}_3)\rangle+  {\rm perm.})\,,
\label{3point}
\label{zeta-3}
\end{eqnarray}
where the `perm.' indicate two other terms with permutations of indices 1, 2 and 3, and whose result we conventionally write as \cite{Chen:2010xk}
\be
\langle \R(\bk_1) \R(\bk_2) \R(\bk_3) \rangle = (2 \pi)^7 \delta (\sum_{i=1}^3 \bk_i) {\cal P}_{\R}^2 \frac{S(k_1,k_2,k_3)}{(k_1 k_2 k_3)^2}
\label{S-def} 
\ee
where ${\cal P}_{\R}$ is the primordial power spectrum given in Eq. \refeq{power-spectrum-R-final} and the dimensionless shape function 
\bea
S=-\frac{1}{2} \invc \frac{1}{ 1-3 \q (3-2\c^2)} \frac{1}{ (1+\T^2)^2}   \left( \Sa+ \T^2 \Ss \right) 
\label{S}
\eea
has been separated out between its purely adiabatic and entropic-induced components, corresponding respectively to the first and the second line in \refeq{zeta-3}. A peculiarity of standard multifield DBI inflation (corresponding to $\tM^2=0$) is that $ \Sa= \Ss$. However, as we will see, this is not the case in multifield DBI Galileon inflation. Hence, we analyse $\Sa$ and $\Ss$ separately in the following, considering first the purely adiabatic signal and then the entropic-induced one.

\subsection{The adiabatic bispectrum}

The adiabatic shape $\Sa$ is given by
\bea
\Sa =3 A_{\dQsi^3} S_{\dQsi^3}+ \frac{\l^2}{2}A_{\dQsi  (\pQsi)^2}  S_{\dQsi (\pQsi)^2}
\label{Sa}
\eea
where $A_{\dQsi^3}$ and $A_{\dQsi  (\pQsi)^2}$ are given in Eqs. \refeq{Afirst} and  \refeq{Asecond} respectively and
\bea
S_{\dQsi^3}&=&\Ao(k_1,k_2,k_3)
\label{A1}
\\
S_{\dQsi (\pQsi)^2} &=& \At(k_1,k_2,k_3)+ 2\,{\rm perm.} \
\label{A2}
\eea
with (this will be useful in our discussion of the entropic-induced shape)
\bea
\Ao(k_1,k_2,k_3)&=&\frac{k_1 k_2 k_3}{K^3} \\
\At(k_1,k_2,k_3)&=&-\frac{k_3}{k_1 k_2 K^3}  \bk_1 \cdot \bk_2 \left(2 k_1 k_2 -k_3 K+   2K^2 \right)\,,
\eea
and $K \equiv  k_1+k_2+k_3 $. The adiabatic shape is thus a linear combination of the well known shapes of $k$-inflationary type $S_{\dQsi^3}$ and $S_{\dQsi (\pQsi)^2}$, which are displayed in Fig. \ref{fig:subfig1}. One can see that there are very similar, in particular both peaking on equilateral triangles $k_1=k_2=k_3$. To assess the experimental ability to distinguish between different bispectrum momentum dependence, we use the scalar product between shapes introduced in \cite{Fergusson:2008ra} (see \cite{Babich:2004gb} for earlier similar definitions). In this sense, the correlation $C[S_{\dQsi^3},S_{\dQsi (\pQsi)^2}]$ equals $0.97$, which explains why the two shapes are often approximated in CMB data analyses by a common (factorizable) ansatz, called equilateral \cite{Creminelli:2005hu}:
\be
S_{{\rm eq}}=-\left(\frac{k_1^2}{k_2 k_3}+2\,{\rm perm.} \right)+\left( \frac{k_1}{k_2}+5\,{\rm perm.} \right)-2\,.
\label{Seq}
\ee
\begin{figure}[h]
\centering
\subfigure[Shape of $S_{\dQsi^3}$ in Eq. \refeq{A1} (left) and $S_{\dQsi (\pQsi)^2}$ in Eq. \refeq{A2} (right).]{
\includegraphics[width=1\textwidth]{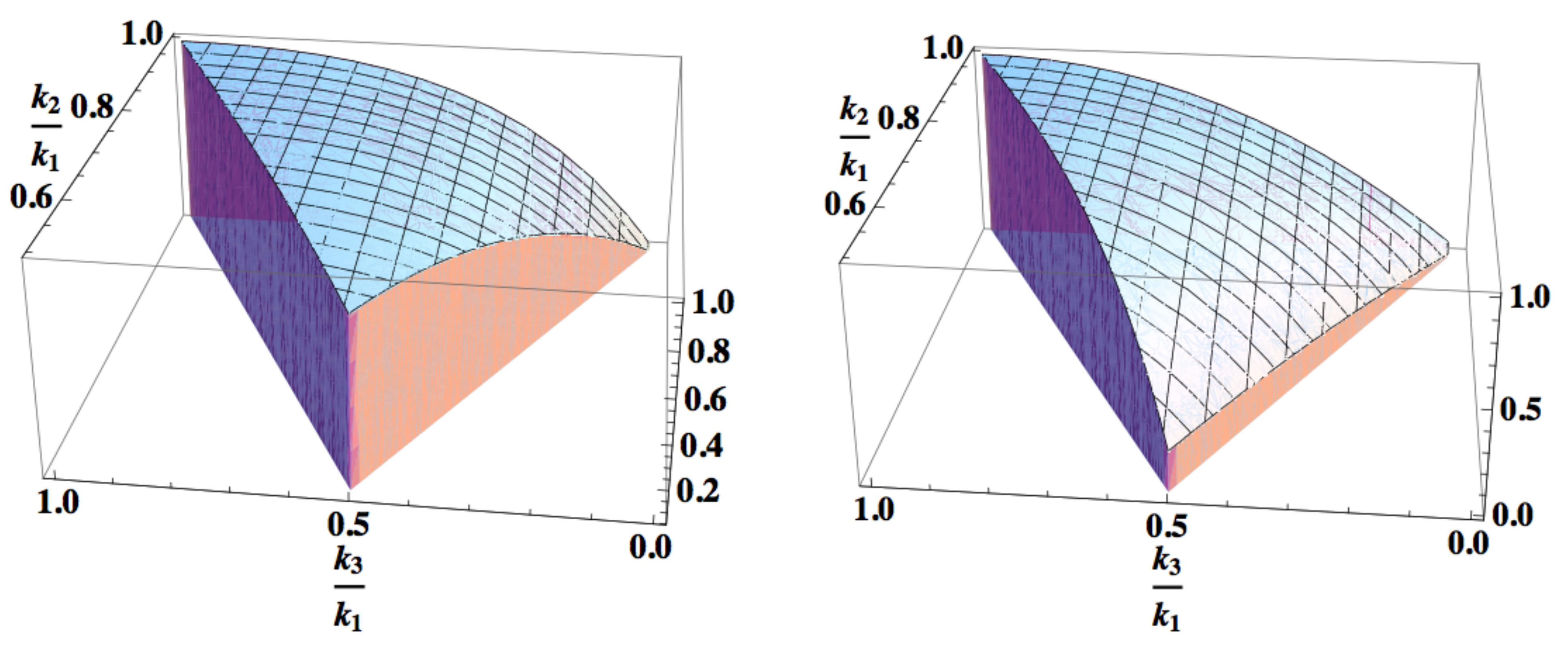}
\label{fig:subfig1}
}
\subfigure[Shape of $S_{{\rm eq}}$ in Eq. \refeq{Seq} (left) and of the absolute value of $S_{{\rm orth}}$ in Eq. \refeq{Sorth} (right).]{
\includegraphics[width=1\textwidth]{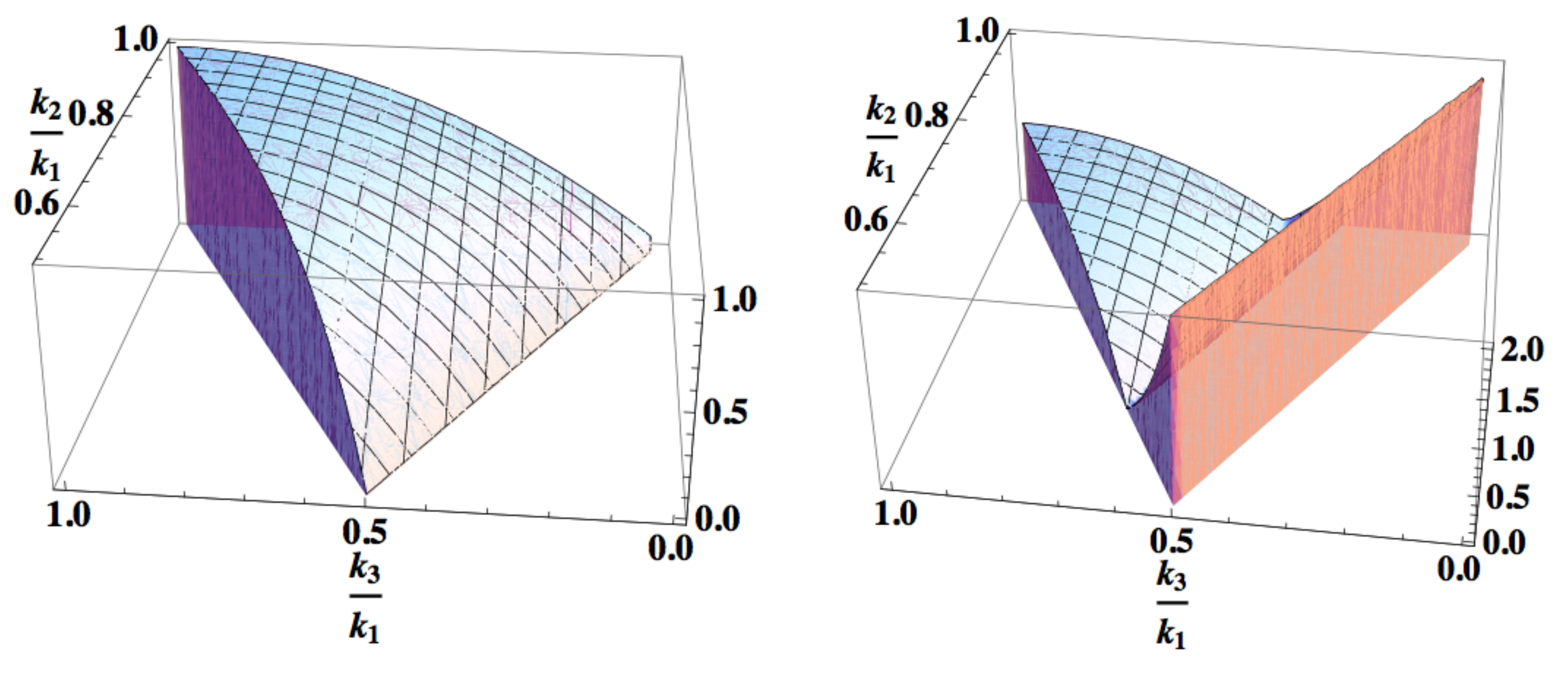}
\label{fig:subfig2}
}
\caption[Optional caption for list of figures]{Different shape $S_X\left(1,\frac{k_2}{k_1},\frac{k_3}{k_1}\right)$ as a  function of $\left(\frac{k_2}{k_1},\frac{k_3}{k_1}\right)$. We set them to zero outside the region $1-k_2/k_1 \leq k_3/k_1 \leq k_2/k_1$ and normalize them to one for equilateral triangles $\frac{k_2}{k_1}=\frac{k_3}{k_1}$.
}
\label{fig:subfigureExample}
\end{figure}
Its shape is displayed in Fig. \ref{fig:subfig2} (left), where one can see that it captures their main features. Indeed, $C[S_{\dQsi^3},S_{{\rm eq}}]=0.94$ and $C[S_{\dQsi (\pQsi)^2},S_{{\rm eq}}]=0.99$.   However, as pointed out in \cite{Chen:2006nt}, $S_{\dQsi^3}$ and $S_{\dQsi (\pQsi)^2}$ are different, especially for squashed triangles $k_2+k_3=k_1/2$ and more generally for flattened triangles $k_2+k_3=k_1$. One can therefore highlight their differences by considering an appropriate linear combination of them with respect to which it is almost orthogonal. This was carried out in Ref. \cite{Senatore:2009gt}, in which the authors designed a corresponding factorizable template, called orthogonal:
\be
S_{{\rm orth}}=-3\left(\frac{k_1^2}{k_2 k_3}+2\,{\rm perm.} \right)+3\left( \frac{k_1}{k_2}+5\,{\rm perm.} \right)-8\,,
\label{Sorth}
\ee
that is represented in Fig. \ref{fig:subfig2} (right) (we actually plot its absolute value so that its difference with the equilateral ansatz for flattened triangles is more visible). This orthogonal shape is now routinely considered in CMB data analyses, which put constraints on its amplitude $f_{NL}^{orth}$ (defined such that $S$ in Eq. \refeq{S-def} equals $\frac{9}{10}f_{NL}^{orth}S_{{\rm orth}}$): $f_{NL}^{orth}=-79.4 \pm 133.3 \,(95 \% \, {\rm C.L})$ (Fergusson et al. \cite{Fergusson:2010dm}) and $f_{NL}^{orth}=-202 \pm 208\, (95 \% \, {\rm C.L})$ (WMAP7 \cite{Komatsu:2010fb})\,.
\begin{figure}[h]
\includegraphics[width=0.5\textwidth]{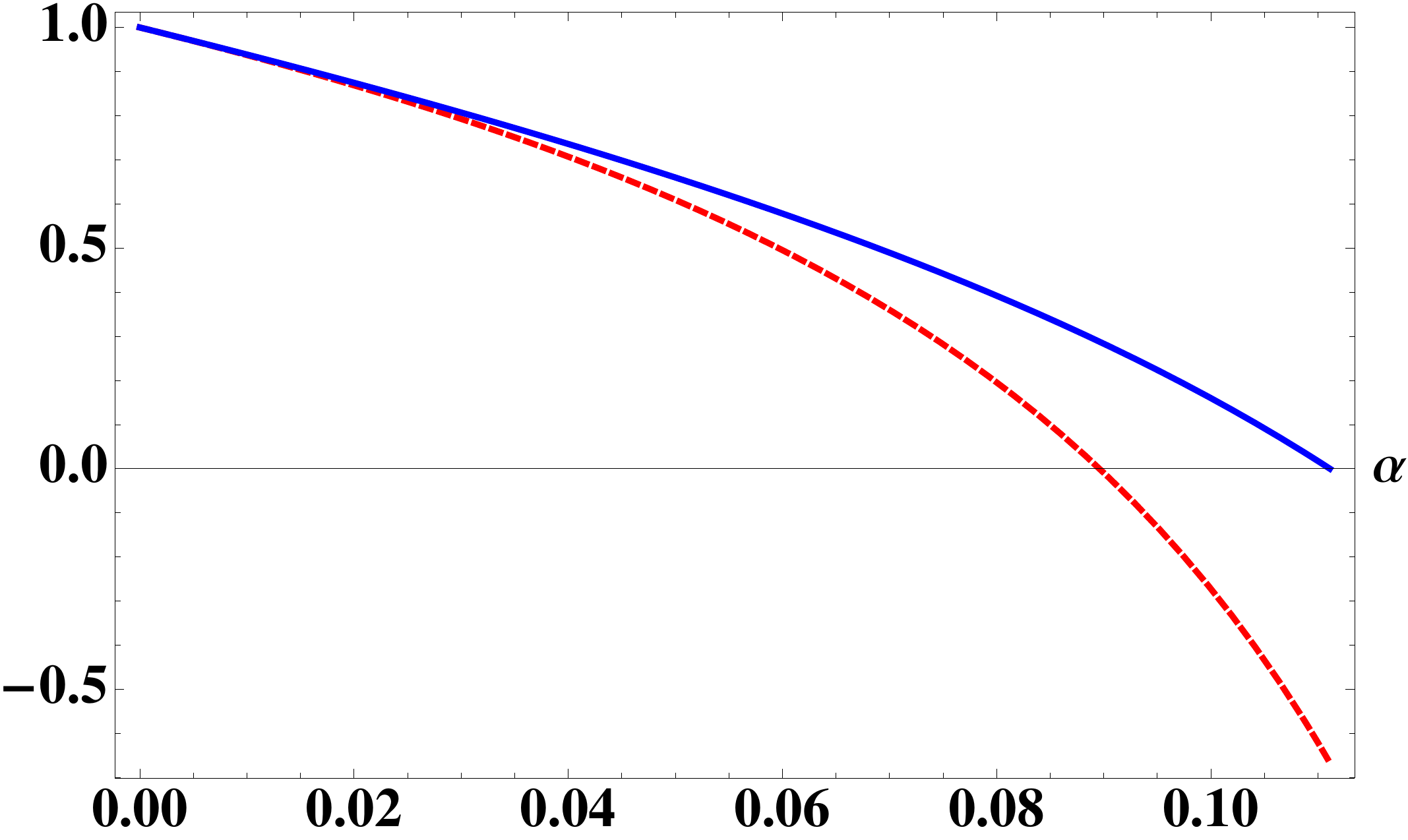}
\caption{\label{As} {\small $A_{\dQsi^3}$ in Eq. \refeq{Afirst} (dashed red) and $A_{\dQsi (\pQsi)^2}$ in Eq. \refeq{Asecond} (plain blue) as we vary $\q$ with $\c^2 \ll 1$.}}
\end{figure}
\begin{figure}[h]
\includegraphics[width=0.5\textwidth]{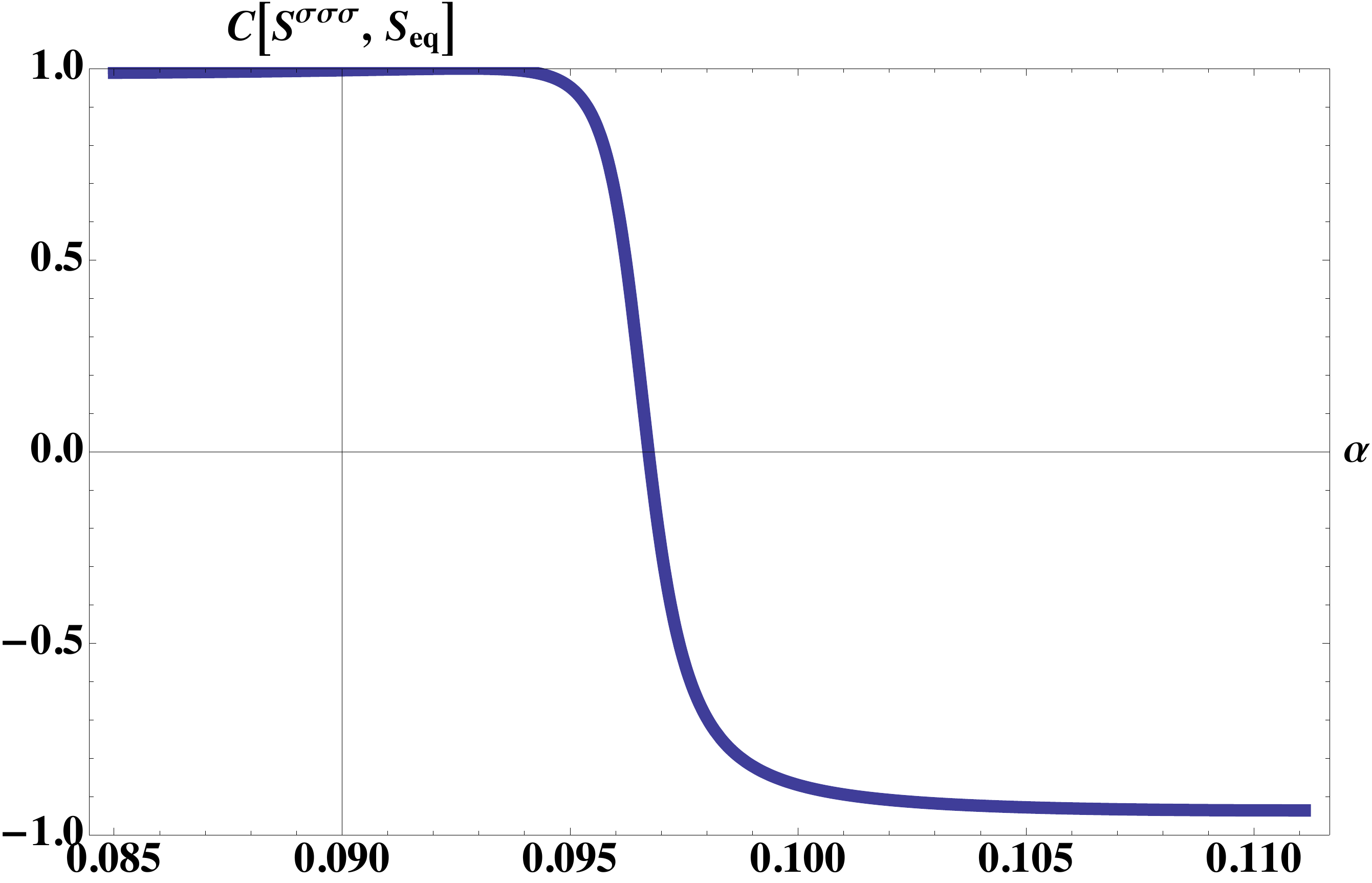}
\space
\space
\includegraphics[width=0.5\textwidth]{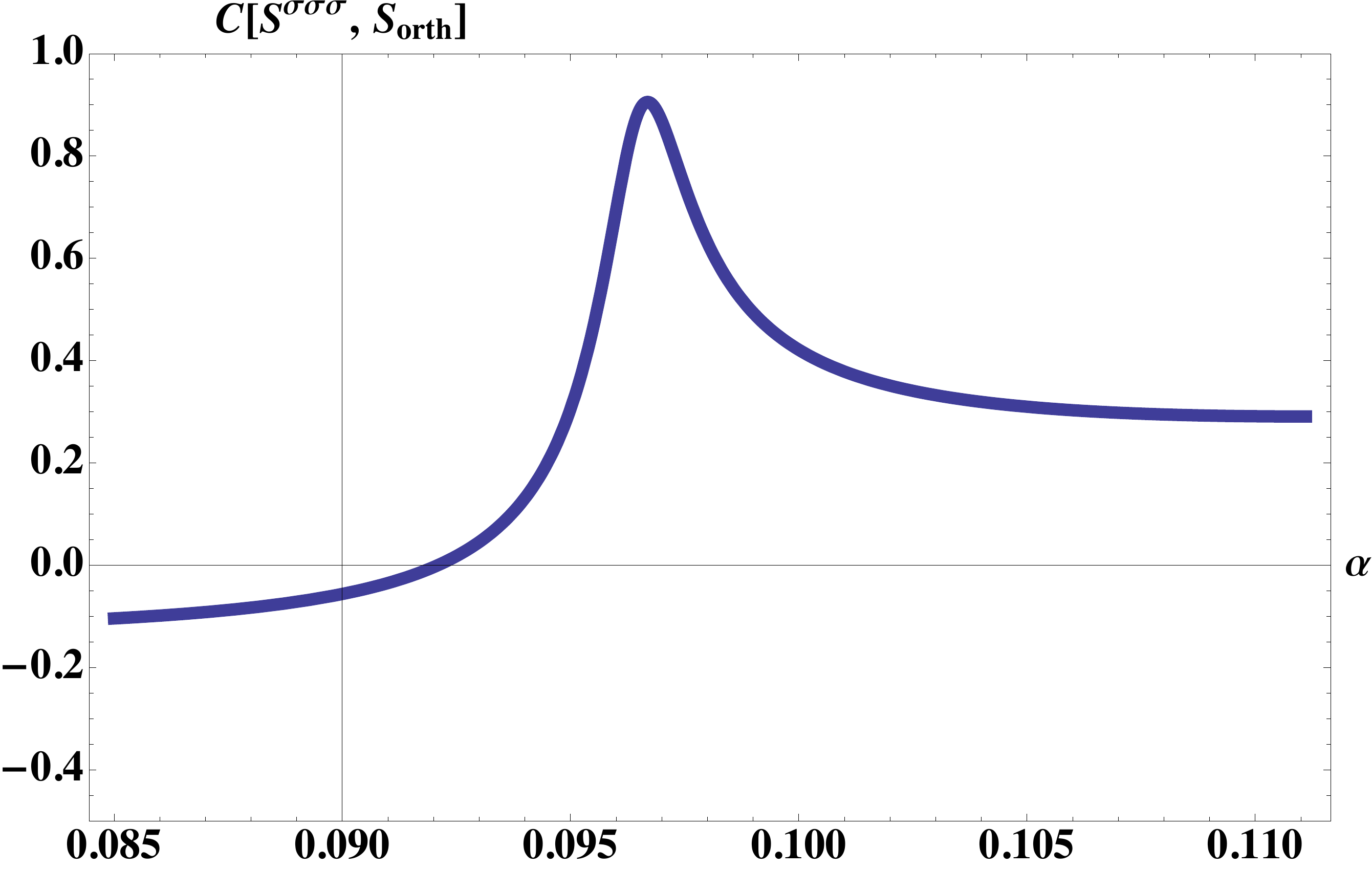}
\caption{\label{correlation} {\small Correlation of the adiabatic shape $S^{(\s \s \s)}$ with the equilateral ansatz (top) and the orthogonal ansatz (bottom) as a function of $\q$ in the relativistic limit $\c^2 \ll 1$.}}
\end{figure}

However, a concrete early universe model that generates such a non-gaussian signal with an observably large amplitude was still lacking\footnote{Shapes referred to as orthogonal in Ref. \cite{Noller:2011hd} reach their maximum in the squashed configuration, like the orthogonal shape. However, their sign is the same for all triangle configurations and their correlation is large not with the orthogonal template but with the equilateral one.}. We now demonstrate that DBI Galileon inflation provides such a model. The reason is simple: as $\q$ increases, $A_{\dQsi^3}$ and $A_{\dQsi (\pQsi)^2}$ deviate from $1$ but, whereas the latter remains positive and tends to $0$, the latter becomes negative (see Fig. \ref{As}). Hence, the adiabatic shape \refeq{Sa}, positively correlated with the equilateral template when $\q=0$, becomes anti-correlated with it as $\q \to 1/9$ (see Fig. \ref{correlation}).
As a consequence, in a transient region -- centered around $\q=0.097$ -- $C[S^{(\s \s \s)},S_{{\rm eq}}]$ becomes small and the similarities between the two equilateral-type shapes are efficiently subtracted, resulting in a large correlation with the orthogonal template of maximum value $91$ \%. One should note that this value is not surprising, as it is the value already given in Ref. \cite{Senatore:2009gt} for the correlation between the true orthogonal shape -- arising as linear combinations of $S_{\dQsi^3}$ and $S_{\dQsi (\pQsi)^2}$ -- and its factorizable template Eq. \refeq{Sorth}. The corresponding shape is represented in Fig. \ref{0.097} (left).
\begin{figure}[h]
\includegraphics[width=1.0\textwidth]{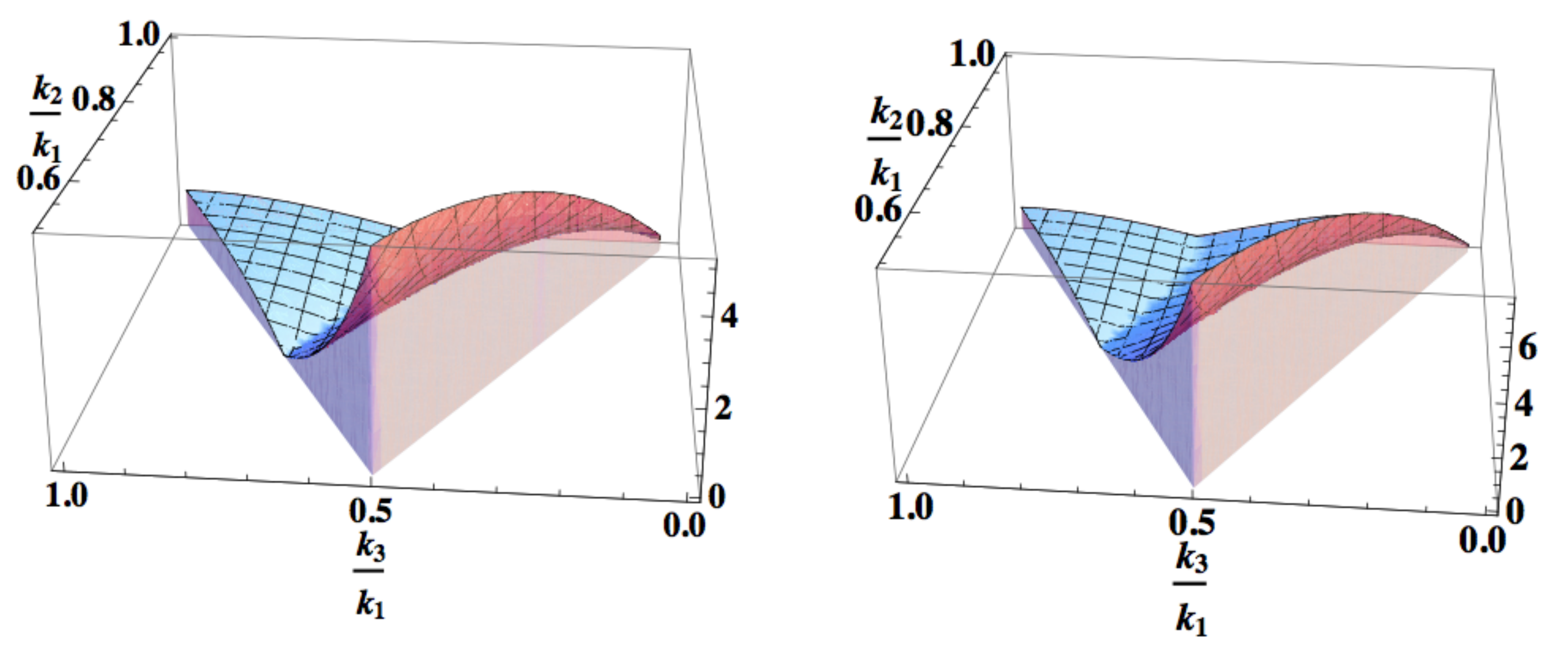}
\caption{\label{0.097} {\small Absolute value of the adiabatic shape $\Sa$ for $\q=0.097$ (left) and of the entropic shape $\Ss$ for $\q=0.080$. We use the same conventions as in Fig. \ref{fig:subfigureExample}.}}
\end{figure}

As far the amplitude of this non-Gaussian signal is concerned, we find that\footnote{This value is calculated as $f_{NL}^{orth}=\frac{10}{9} \frac{C[S,S_{{\rm orth}}]}{C[S_{{\rm orth}},S_{{\rm orth}}]}$, as is relevant to compare to the observational constraints \cite{Fergusson:2010dm,Komatsu:2010fb}. Alternatively, one can estimate it by simply evaluating $\frac{10}{9} S$ on equilateral triangles. One then finds $f_{NL}^{orth\, (\s \s \s)}\simeq -\frac{0.011}{\c^2 (1+\T^2)^2}$.} $f_{NL}^{orth \,(\s \s \s)}=-\frac{0.016}{\c^2 (1+\T^2)^2}$ (at $\q=0.097$). Outside the transient region -- for definiteness let us say for $\q < 0.095$ and $\q > 0.100$, the adiabatic shape can be considered as truly equilateral, and its amplitude simply measured by evaluating $\frac{10}{9} S$ on equilateral triangles. One then finds
\begin{eqnarray}
f_{\rm NL} ^{eq \,(\s\s\s)} =-\frac{5}{324 \c^2} \frac{1}{(1+\T^2)^2}
\frac{  \left(
  21-404 \q +2233 \q^2 -3066\q^3\right)}{\left(1-5 \q  \right)^2 \left(1-9\q \right)}\,,
\label{fnl_equil_ad_prediction}
\end{eqnarray}
whose dependence on $\q$ is represented in Fig. \ref{Amplitude-ad} (left).
\begin{figure}[h]
\includegraphics[width=1.0\textwidth]{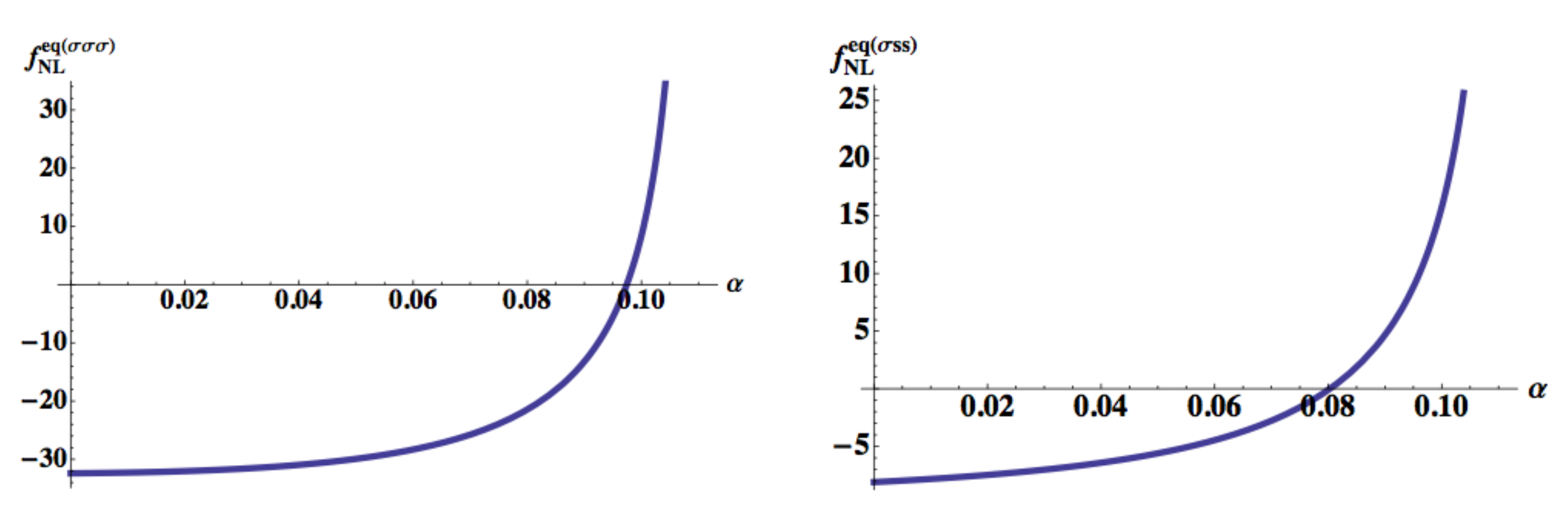}
\caption{\label{Amplitude-ad} {\small $f_{\rm NL} ^{eq \,(\s\s\s)}$ in Eq. \refeq{fnl_equil_ad_prediction} as a function of $\q$ for $\c=0.1$ and $\T^2=0$ (left). $f_{\rm NL} ^{eq \,(\s s s)}$ in Eq. \refeq{fnl_equil_en_prediction} as a function of $\q$ for $\c=0.1$ and $\T^2=1$ (right).}}
\end{figure}
Consistently with the correlation of the adiabatic shape with the equilateral ansatz (top of Fig. \ref{correlation}), one finds that, as $\q$ increases, $f_{\rm NL} ^{eq \,(\s\s\s)}$ goes from negative to positive values. It also grows unboundedly as $\q \to 1/9$, corresponding to the fact that the kinetic term of the adiabatic fluctuation becomes close to zero and that the theory becomes strongly coupled in this regime. To be more quantitative, the figure \ref{Allowed} (left) gives the region allowed by the current observational bounds on non-Gaussianities in the two-dimensional parameter space $(\c,\q)$ for $\T=0$. We used the constraints $f_{\rm NL}^{eq}=143.5 \pm 151.2$ and $f_{\rm NL}^{orth} = -79.4 \pm 133.3$ $(95 \% \, {\rm C.L})$ from Fergusson and Shellard \cite{Fergusson:2010dm} and took into account the condition that the shape is sufficiently correlated (or anti-correlated) at $80 \%$ with the equilateral or orthogonal templates. The unbounded growth of $f_{\rm NL} ^{eq \,(\s\s\s)}$ with $\q$ of course explains the sharpness of the constraints when $\q \gtrsim 0.10$. Note also that smaller values of $\c$ are allowed for $\q \simeq 0.097$ comparatively to other values of $\q$ because of the small numerical factor in $f_{NL}^{orth}=-\frac{0.016}{\c^2 (1+\T^2)^2}$, which itself is explained by the partial cancellations between $S_{\dQsi^3}$ and $S_{\dQsi (\pQsi)^2}$ required to generate orthogonal non-Gaussianities.\\
\begin{figure}[h]
\includegraphics[width=1.0\textwidth]{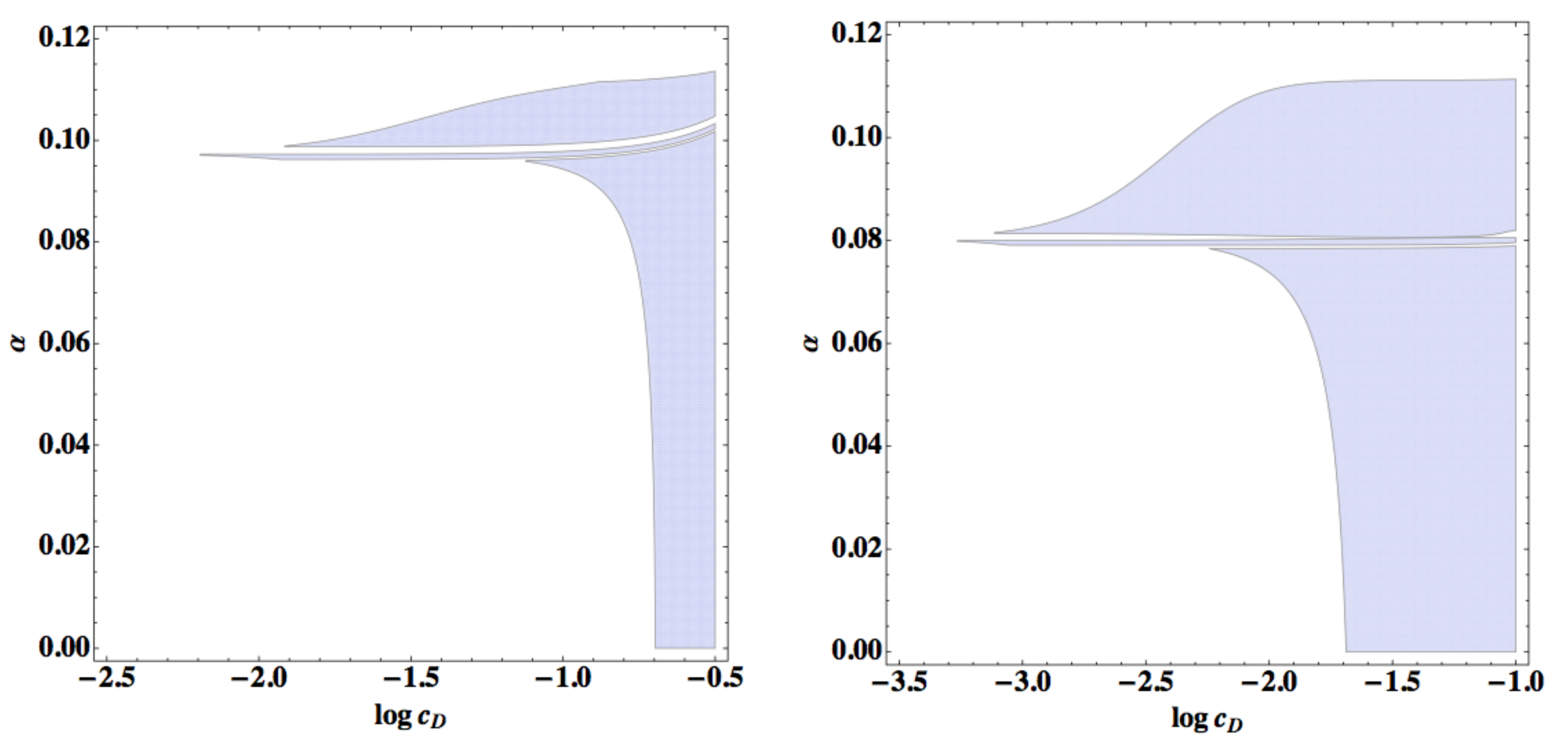}
\caption{\label{Allowed} {\small  The region in the parameter space $(\c,\q)$ allowed by the observational constraints on $f_{\rm NL}^{eq}$ and $f_{\rm NL}^{orth}$ for $\T=0$ (left) and $\T=10$ (right).}}
\end{figure}
\begin{figure}[h]
\includegraphics[width=0.6\textwidth]{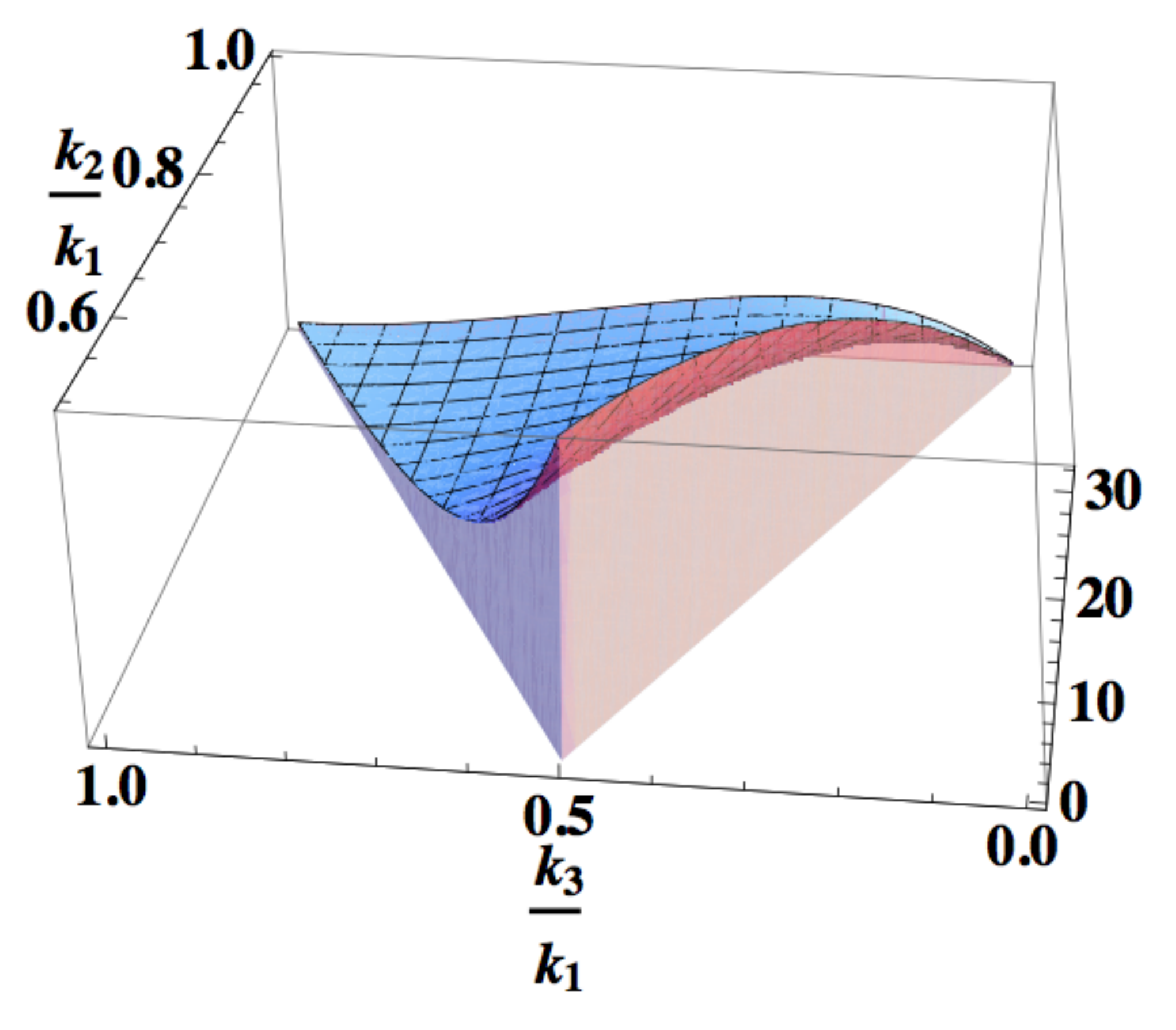}
\caption{\label{Surfing} {\small The adiabatic shape $\Sa$ for $\q=0.0975$. We use the same conventions as in Fig. \ref{fig:subfigureExample}.}}
\end{figure}
We have concentrated so far on the equilateral and orthogonal shapes of non-Gaussianities. However, one should point out that shapes yet qualitatively different from them do arise in our model. In particular, as is clear from fig. \ref{correlation}, neither the equilateral nor the orthogonal templates faithfully represent the adiabatic shape at $\q \simeq 0.0975$ (the corresponding correlations being respectively $-0.55$ and $0.74$). Indeed, the corresponding shape, represented in Fig. \ref{Surfing}, is very close to what the authors of Ref. \cite{Senatore:2009gt} have called the flat shape. To be complete, one should actually note that shapes very similar were shown to arise in different contexts: the ``surfing shape'' of Ref. \cite{Creminelli:2010qf} or the ``flat'' shapes generated by curvature-related terms in \cite{Bartolo:2010bj}. Eventually, all of those shapes also have a large overlap with the enfolded template \cite{Meerburg:2009ys}, aimed at modeling the effect of a non-Bunch Davies vacuum \cite{Chen:2006nt,Holman:2007na}, and which is actually a linear combination of the equilateral and orthogonal template \cite{Senatore:2009gt}. For instance, our shape in Fig. \ref{Surfing} is 93 $\%$ correlated with it.

\subsection{The entropic-induced bispectrum}

The entropic-induced shape $\Ss$ in Eq. \refeq{S} reads
\be
\Ss =\chi  \left(\l^2 S_{\dQsi \kineticperp}-\frac12 S_{ \dQsi (\pQs)^2}+\frac{\l^2}{2}(1-2\q) S_{ \pQsi \pQs \dQs}
-(1-\l^2) \frac{\q}{2} \left(  3  \l^2 S_{ \kineticperp \LapQsi}-S_{ \gradperp  \LapQsi}   \right)
   \right)
\label{Sentropic}
\ee
with
\be
\chi \equiv \frac{1}{1-3\q}\sqrt{(1-3\q(3-2\c^2))(1-\q (5-2\c^2))}
\ee
and
\bea
S_{\dQsi \kineticperp}&=& \frac{1}{\l^2} \Ao(k_1,\l k_2,\l k_3)+2\,{\rm perm.} 
\label{As1s} \\
S_{ \dQsi (\pQs)^2}&=&  \At(\l k_2,\l k_3,k_1)+2\, {\rm perm.} 
\label{As21s}\\
S_{ \pQsi \pQs \dQs}&=& \frac{1}{\l}\At(k_1,\l k_2, \l k_3)   + 5\, {\rm perm.}
\label{As22s}\\
S_{ \kineticperp \LapQsi}&=&   \frac{k_1 k_2 k_3}{\Kone^4} \left(\Kone+3 k_1 \right)+2\, {\rm perm.}
\label{As41s}
\eea
\bea
&& \hspace*{-3.0em}S_{ \gradperp  \LapQsi}=  \frac{k_1}{k_2 k_3} \frac{\bk_2 \cdot \bk_3}{\Kone^4}   \left(\Kone^3+k_1 \Kone^2-(k_1^2-\l^2 k_2 k_3) \Kone+3 \l^2 k_1 k_2 k_3  \right)  +2\, {\rm perm.} 
\label{As31s}
\eea
with $\Kone=k_1+\l (k_2+k_3)$. The fact that $\chi$ verifies $0 < \chi \leq 1$ with $\chi=1$ when $\q=0$ implies that the induced gravity tends to lower the intrinsic amplitude of quantum non-Gaussianities induced by the entropic perturbations compared to the ones generated by the adiabatic perturbations. However, depending on the magnitude of the transfer function $\T$, and hence on the large-scale evolution, the total bispectrum can well originate mostly from entropic origin.

Similarly to the purely adiabatic sector, in the relativistic regime $\c^2 \ll 1$, the entropic-induced shape $\Ss$ represents a-one parameter family of shapes depending on $\q$. However, it is also interesting to study each of the shapes \refeq{As1s}-\refeq{As31s} separately\footnote{There is some arbitrariness here because, as we have pointed out, some vertices, and hence some shapes, may well have been replaced by others by using the linear equations of motion and integrating by parts.}. One should note then that one can not factor out $\lambda$ from their $k_i$'s dependence. In other words, these shapes constitute themselves five families of $\lambda$-dependent shapes, which one can check to be linearly independent. In this respect, $\lambda=1$, corresponding to standard multifield DBI inflation in which the adiabatic and the entropic speed of sound coincide, is an enhanced symmetry point at which the entropic-induced shapes can be expressed in terms of the two purely adiabatic ones \refeq{A1} and \refeq{A2}:
\bea
 \underline{\l=1} \qquad \qquad  \frac13 S_{\dQsi \kineticperp}&=& \frac16 S_{ \kineticperp \LapQsi}=S_{\dQsi^3} \\
S_{ \dQsi (\pQs)^2}&=&\frac12 S_{ \pQsi \pQs \dQs}=S_{\dQsi (\pQsi)^2} \\
S_{ \gradperp  \LapQsi}&=&6 S_{\dQsi^3}-S_{\dQsi (\pQsi)^2}\,.
\eea
However, for general values of $\lambda$, some of the entropic-induced shapes differ significantly from the equilateral template. The most dramatic changes occur for $S_{\dQsi \kineticperp}$ and $S_{ \kineticperp \LapQsi}$, which are represented in Fig. \ref{Local} for $\lambda=0.1$.
\begin{figure}[h]
\includegraphics[width=1.0\textwidth]{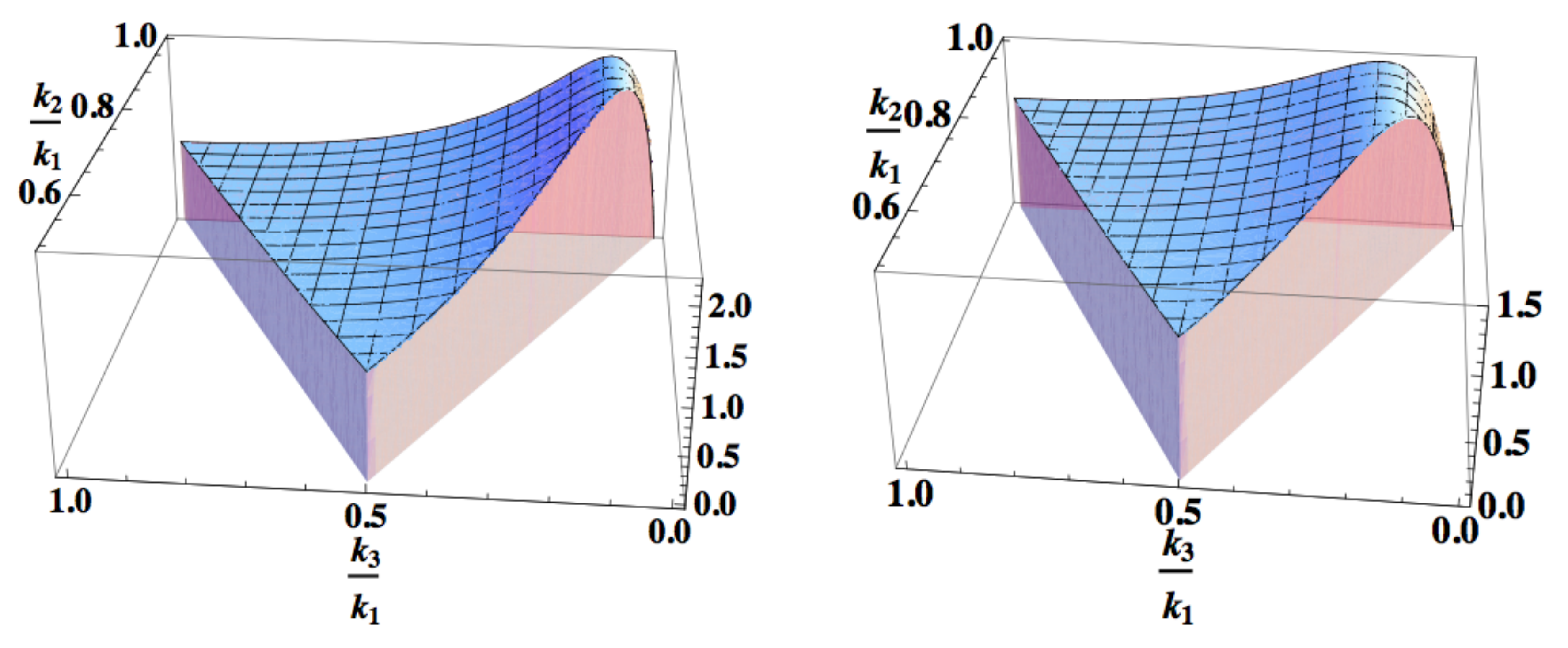}
\caption{\label{Local} {\small Shape of $S_{\dQsi \kineticperp}$ in Eq. \refeq{As1s} (left) and $S_{ \kineticperp \LapQsi}$ in Eq. \refeq{As41s} (right) for $\lambda=0.1$. We use the same conventions as in Fig. \ref{fig:subfigureExample}.}}
\end{figure}
One can see that these shapes reach their maximum near the squeezed limit $k_3 \ll k_2 \simeq k_1$ (all the more so as $\lambda$ decreases), which is unusual for non-Gaussianities generated around horizon crossing\footnote{One should note however that, as expected, they vanish as $k_3/k_1$ in the squeezed limit, like all the other shapes generated in this model.}. Interestingly, note that the authors of Ref. \cite{Burrage:2011hd} have recently shown that similar shapes (referred to as $S'_{\eps}$ and $S_{s}$ in their paper) arise at next-to-leading order in slow-roll in $k$-inflation. Now coming back to the global entropic-induced shape $\Ss$, one can see from its expression \refeq{Sentropic} that the contributions from the shapes $S_{\dQsi \kineticperp}$ and $S_{ \kineticperp \LapQsi}$ are diminished by factors of $\l^2$ compared to others. Hence, we do not expect the total entropic-induced shape $\Ss$ to exhibit features similar to the ones in Fig. \ref{Local}. Indeed, representing it graphically shows that it is qualitatively similar to the adiabatic one: it interpolates between negative (when $\q=0$) and positive equilateral non-Gaussianities passing through orthogonal ones, although for smaller values of $\q$ centered around $0.08$ (the corresponding shape is represented on the right of Fig. \ref{0.097}). This is confirmed by the correlations of $\Ss$ with the equilateral and orthogonal templates, shown as a function of $\q$ in Fig. \ref{Correlation-entropic}.
\begin{figure}[h]
\includegraphics[width=0.5\textwidth]{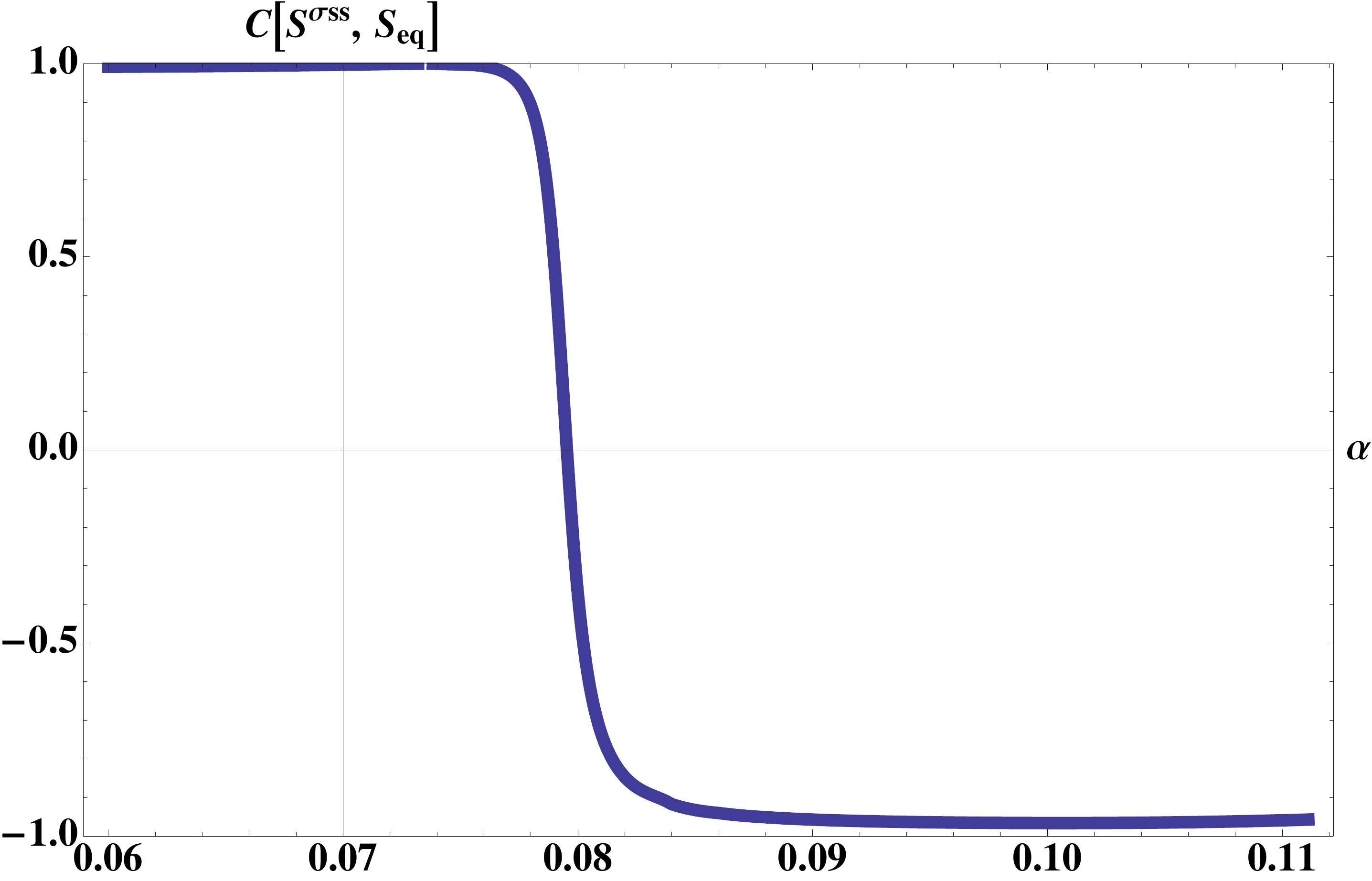}
\space
\space
\includegraphics[width=0.5\textwidth]{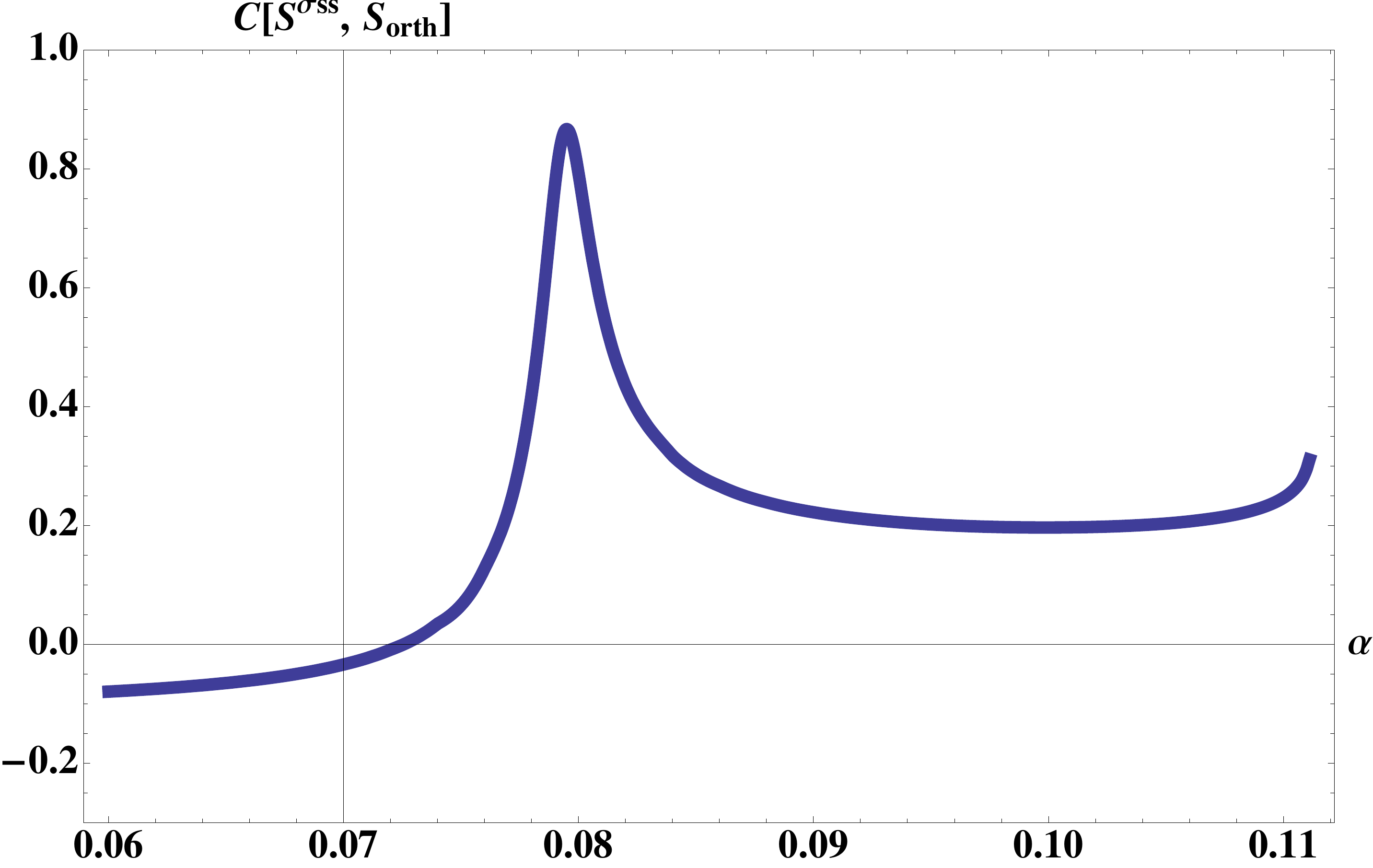}
\caption{\label{Correlation-entropic} {\small Correlation of the entropic-induced shape $S^{(\s s s)}$ with the equilateral ansatz (top) and the orthogonal ansatz (bottom) as a function of $\q$ in the relativistic limit $\c^2 \ll 1$.}}
\end{figure}
Let us note also that $\Ss$ is of ``surfing'' type, similar to the adiabatic shape in Fig. \ref{Surfing}, for values of $\q$ slightly larger than the transition value $0.08$. 

Regarding the amplitude of the entropic-induced non-Gaussianities, we find that $f_{\rm NL} ^{orth \,(\s ss)} =-\frac{0.0043}{\c^2}\frac{\T^2}{(1+\T^2)^2 }$ for $\q=0.08$ while it can be simply characterized by evaluating $\frac{10}{9}S$ on equilateral triangles for $\q < 0.078$ and $\q > 0.082$, leading to
\bea
&&f_{\rm NL} ^{eq (\s ss)} =-\frac{\T^2}{(1+\T^2)^2 }  \frac{5}{12 (1+2 \lambda)^4 \c^2}
\frac{\sqrt{(1-9\q)(1-5\q)}  }{(1-3\q)  \left(1-9 \q \right) (1-5\q)^2}  \nn\\
 \hspace*{-5.0em} && \times \left( 33+30 \l-\q (762+648\l)+3 \q^2 (1631+1282 \l)-\q^3(6532+5276 \l) \right)\,,
\label{fnl_equil_en_prediction}
\eea
where $\l \simeq \sqrt{(1-9\q)/(1-5\q)}$ in the relativistic regime. The dependence of $f_{\rm NL} ^{eq (\s ss)}$ on $\q$, similar to the one of $f_{\rm NL} ^{eq \,(\s\s\s)}$ in Eq. \refeq{fnl_equil_ad_prediction}, is displayed in Fig. \ref{Amplitude-ad} (right). The observational constraints on the parameter space $(\c,\q)$ for $\T=10$ -- in which case the total bispectrum is dominated by its entropic-induced component -- can be seen in Fig. \ref{Allowed} (right). Values of $\c$ smaller than in the case of purely adiabatic non-Gaussianities are allowed because of the suppression factor $\frac{\T^2}{(1+\T^2)^2}$ but the pattern of constraints is qualitatively similar.\\

In all our above discussion on the non-Gaussianities generated by multifield DBI Galileon inflation, we considered the relativistic regime $\c^2 \ll 1$ in which large non-Gaussianities are generated. As we have explained, for bigger values if $\c^2$, the gravitational back-reaction can not be neglected and our analysis is not complete. However, to get a taste of the shape of non-Gaussianities for intermediate values of $\c^2$, it is interesting to take the non-relativistic limit $\c \to 1$ (and hence $\l \simeq 1$) in our expression \refeq{S} for the shape. The overall amplitude of non-Gaussianities become of course small in this limit (as $1-\c^2$) but interestingly, the corresponding shape simplifies to
\bea
\lim_{\c \to 1 }  \frac{S(k_1,k_2,k_3)}{1-\c^2} &=&- \frac{1}{2(1-3 \q)}  \frac{1}{(1+\T^2)}    \left( 3  S_{\dQsi^3}+ \left(\frac12 -2 \q \right) S_{\dQsi (\pQsi)^2}  \right)\,.
\label{NR}
\eea
In particular, the entropic-induced shape and the adiabatic shape become equal in the non-relativistic limit, so that the total shape becomes independent of the transfer function $\T$, as in standard multifield DBI inflation. Note also that, as $\frac12 -2 \q > 0$, there is no cancellation between $S_{\dQsi^3}$ and $S_{\dQsi (\pQsi)^2}$ in Eq. \refeq{NR}, so that this shape is of equilateral type for any values of the induced gravity strength $\q$. Of course, these results should be taken with a grain of salt as other terms that we have neglected compete in magnitude with the result \refeq{NR} above. Nonetheless, they give an indication that the rich structure of shapes that we have studied in this paper does not persist in the non-relativistic regime.

\section{C\lowercase{onclusions}}
\label{Conclusion}

In this paper we have studied the cosmological implications in the early universe of the multifield relativistic Galileon model in which an induced gravity term is added to the DBI action governing the motion of a D3-brane in a higher-dimensional spacetime.  By employing a bimetric perspective in which the induced gravity and Einstein-Hilbert action are treated on equal footing, we have been able to analyze this model in a physically transparent and computationally very efficient way. After deriving the gravitational and field equations of motion in full generality, we have considered the background homogenous evolution. We have shown that the induced gravity tends to violate the null energy condition and to hamper inflation in the relativistic regime in which the brane almost saturates its speed limit. Although the induced gravity should thus be subdominant for our model to sustain a phase of quasi de-Sitter expansion in the relativistic regime, we have shown that its effects need not be negligible. 

Our study of the cosmological perturbations generated during inflation even revealed that the induced gravity can leave very non-trivial imprints on the cosmological fluctuations. We computed the exact second-order action in the fluctuations, showing the existence of a critical background energy density above which cosmological fluctuations become ghosts. We have also shown how the fluctuations react both to the cosmological and to the induced metric as a consequence of the bimetric structure of our model. Restricting our attention to the case of two fields, we were able to cast the exact coupled linear equations of motion for the adiabatic and entropic perturbations in a very elegant and intuitive form. An interesting feature of our model lies in the fact that neither these perturbations not the field fluctuations themselves are decoupled on sub-horizon scales. Leaving the study of its consequences to future works, we neglected this coupling and derived the properties of the cosmological fluctuations at leading order in a slow-varying approximation. We then showed that the entropic speed of sound differ from (and is less than) the adiabatic one. As a consequence, the two types of perturbations are amplified at their respective sound-horizon crossing and the amplitude of entropic perturbations are even more enhanced, compared to standard multifield DBI inflation, than the adiabatic ones.

We then considered the non-Gaussianities of the curvature perturbation generated in the relativistic regime. For that purpose, we computed the dominant contributions to the third-order action and we identified its seven independent cubic operators: two purely adiabatic and five mixed adiabatic/entropic ones. As a consequence, the shape of the primordial bispectrum induced by the entropic fluctuations is different from the purely adiabatic shape, thereby breaking the degeneracy of standard multifield DBI inflation. Each of these components of the bispectrum depends non-trivially on the strength of the induced gravity as measured by a parameter $\q <1/9$. Although quantitatively different, they share common features: their shape is of equilateral-type for small and large (the largest allowed) values of $\q$, $f_{{\rm NL}}^{eq}$ being then respectively negative and positive. In transient regions, centered respectively around $\q=0.097$ and $\q=0.08$, the two shapes become of orthogonal type. We stress that it is the first concrete model in which this shape of non-Gaussianities, which is routinely considered in CMB data analyses, is generated with an observably large amplitude. More generally, we find quite remarkable the fact that qualitative aspects of the bispectrum, such as its shape and sign, depend very sensitively on the precise value of the induced gravity strength. This exemplifies the usefulness of higher-order statistics for pinning down the mechanism that seeded the large-scale structure of the Universe.

Eventually, the fact that the induced gravity tends to violate the null energy condition is potentially interesting in the context of bouncing cosmologies, in which this condition should be broken in order to connect a contracting to an expanding universe. We leave for future studies the interesting question of whether a stable bounce solution can be constructed in DBI Galileon models.

\medskip

\begin{acknowledgments}

We would like to thank X. Chen, C. Deffayet, A. De Fellice, D. Langlois, E. Lim, L. McAllister, J. Noller, F. Piazza, C. de Rham, R. Ribeiro, D. Seery, S. Tsujikawa, D. Wands and M. Yamaguchi for interesting discussions related to the topic of this paper. SRP is supported by STFC and the Centre for Theoretical Cosmology, SM is supported by JSPS and KK is supported by STFC grant ST/H002774/1, the ERC and the Leverhulme trust.

\end{acknowledgments}

\appendix

\section{Details on the explicit form of the action}
\label{R[q]explicit}

In this appendix, we give the explicit form of the induced gravity action $S_{{\rm ind}}=\frac{\e}{2} \int  {\rm d}^4 x \sqrt{-\g} R[\g]$ in terms of the scalar fields and the geometrical quantities associated to the cosmological metric, leading to a multifield relativistic extension of the quartic Galileon Lagrangian in curved spacetime.

Remembering that $\g_{\mu \nu}=h^{-1/2} \tg_{\mu \nu}$ where $\tg_{\mu \nu} \equiv  g_{\mu \nu}  + f \, G_{IJ} \partial_\mu \phi^I \partial_\nu \phi^J$, one readily obtains
\be
S_{{\rm ind}}= \int {\rm d}^4 x  \frac{\tM^2}{2} \sqrt{-\tg} R[\tg]- \int {\rm d}^4 x \sqrt{-\tg} \frac{3 \tM^2}{8} \frac{f_{,I} f_{,J}}{f^2}X^{IJ}_{\tg}
\ee
where $X^{IJ}_{\tg} \equiv -\frac12 \tg^{\mu \nu}\partial_{\mu} \phi^I \partial_{\mu} \phi^J$ and $\tM^2 \equiv M^2/\sqrt{h}$. The real difficulty is to obtain an explicit expression of the inverse $\tg^{\mu \nu}$ of $\tg_{\mu \nu}$ as well of its Ricci scalar, the two metrics $\tg_{\mu \nu}$ and $g_{\mu \nu}$ being related by a disformal transformation. Using the results in Ref. \cite{Silva:fk}, it turns out one  can obtain the following non-trivial expression for $\tg^{\mu \nu}$:
\be
\tg^{\mu \nu}=\a g^{\mu \nu}-\A_{IJ} \nabla^{\mu} \phi^I \nabla^{\nu} \phi^J 
\ee
where
\be
\a \equiv \frac{1}{\cal D} \left(1-2fX^{I}_I+4f^2 X^{[I}_IX_J^{J]} -8f^3X^{[I}_IX_J^{J} X_K^{K]} \right) 
\ee
and
\be
\A_{IJ} \equiv  \frac{f}{\cal D} \left( \left(1-2fX^{A}_A+4f^2 X^{[B}_BX_C^{C]}\right)G_{IJ}+2f\left(1-2fX_K^K\right)X_{IJ}+4f^2X_{IK}X^{K}_J \right) \,.
\ee
Here and in the following, field space indices $I,J \ldots$ are lowered (respectively raised) with the field space metric $G_{IJ}$ (respectively its inverse). $\D$ is given explicitly in Eq. \refeq{def_explicit} in terms of the kinetic terms $X^{IJ}$ defined in Eq. \refeq{def-XIJ} and $\nabla^{\mu}$ denotes the covariant derivative associated to the cosmological metric $g_{\mu \nu}$. Note also that all this complexity disappears in the single field case, in which $A \to 1$, $\A_{IJ} \to f/(1-2f X)$ and $\D \to 1-2f X$. 

To calculate $R[\tg]$, one possibility is to use the formulas relating different covariant derivatives and their related geometrical quantities, as given for example in Ref. \cite{Wald}. Here, we rather follow a more down-to-Earth approach. We start from the general expression
 \be
 R_{\lambda \mu \nu \kappa}[\tg]=  \half \left(\tg_{\mu \nu,\kappa \lambda}+\tg_{\lambda \kappa,\mu \nu}-\tg_{\lambda \nu,\kappa \mu} -\tg_{\mu \kappa,\nu \lambda}\right) +\tg_{\eta \sigma} (\Gamma[\tg]^{\eta}_{\kappa \lambda} \Gamma[\tg]^{\sigma}_{\mu \nu}-\Gamma[\tg]^{\eta}_{\nu \lambda} \Gamma[\tg]^{\sigma}_{\mu \kappa} ) 
 \ee
that we evaluate in a locally inertial frame (in which $\partial_{\gamma} g_{\alpha \beta}=0$ and hence $\Gamma[g]^{\alpha}_{\beta \gamma}=0$) before covariantising it. In such a frame, one finds that
 \be
 \Gamma[\tg]^{\alpha}_{\beta \gamma}= f( \a  G_{IJ}    +  2 \A_{IK} G_{JM} X^{KM})   \nabla^{\alpha} \phi^{I} ({\partial}_{\beta} \partial_{\gamma} \phi^J+\hat{\Gamma}^J_{AB}\nabla_{\beta} \phi^A \nabla_{\gamma} \phi^B)
 \ee
where a hat denotes a quantity defined with respect to the metric $\hat{G}_{IJ} \equiv f G_{IJ}$, so that
\be
 \hat{\Gamma}^J_{AB}=\frac{1}{2 f}(\delta^J_A f_{,B} +\delta^J_B f_{,A}-G^{JI} f_{,I}G_{AB})+\frac12 G^{JC}(G_{CA,B}+G_{CB,A}-G_{AB,C})\,.
 \label{total-Gamma}
 \ee
 In this frame, one also has
 \bea
 \tg_{\mu \nu,\kappa \lambda}+\tg_{\lambda \kappa,\mu \nu}-\tg_{\lambda \nu,\kappa \mu} -\tg_{\mu \kappa,\nu \lambda}&=&g_{\mu \nu,\kappa \lambda}+g_{\lambda \kappa,\mu \nu}-g_{\lambda \nu,\kappa \mu} -g_{\mu \kappa,\nu \lambda} \nonumber \\
 &-&4fG_{IJ}(\phi^I_{,[\overline{\lambda} \kappa}+\hat{\Gamma}^I_{AB}\phi^A_{,[\overline{\lambda}} \phi^B_{,\kappa})   (\phi^J_{,\overline{\mu}] \nu}+\hat{\Gamma}^J_{CD}\phi^C_{,\overline{\mu}]} \phi^D_{,\nu})  \nonumber \\
 &+&  2 \hat{R}_{LIJK}\phi^K_{,\kappa} \phi^L_{,\lambda} \phi^I_{,\mu} \phi^J_{,\nu}
 \eea
and
\bea
\tg_{\eta \sigma} (\Gamma[\tg]^{\eta}_{\kappa \lambda} \Gamma[\tg]^{\sigma}_{\mu \nu}-\Gamma[\tg]^{\eta}_{\nu \lambda} \Gamma[\tg]^{\sigma}_{\mu \kappa}  \Gamma[q]) &=&-2(\phi^A_{,[\overline{\lambda} \kappa}+\hat{\Gamma}^A_{IJ}\phi^I_{,[\overline{\lambda}} \phi^I_{,\kappa}) 
 (\phi^B_{,\overline{\mu}] \nu}+\hat{\Gamma}^B_{CD}\phi^C_{,\overline{\mu}]} \phi^D_{,\nu}) 
  (H_{AB}-f G_{AB}) \nonumber
\eea
with 
\bea
\H_{AB} &\equiv& f G_{AB}+2 \a^2 f^2 (X_{AB}-2fX_{AC}X_{B}^C) +8 \a f^2 X_{(\underline{A}C} \A^{E C}(X_{\underline{B})E}-2fX_{\underline{B})}^H X_{EH}) \nn \\
&+& 8 f^2 \A^{GC} \A^{EF} X_{AC} X_{BF} (X_{EG}-2f X_{EH} X_G^{H} )\,.
\eea
Covariantizing the result, i.e., with the substitution $\phi^I_{,\mu \nu} \to \nabla_\mu \nabla_\nu \phi^I$, one deduces the compact expression
\bea
R_{\lambda \alpha \nu \beta}[\tg]=R_{\lambda \alpha \nu \beta}[g]-H_{AB}(\hP_{\lambda \beta}^A \hP_{\alpha \nu}^B-\hP_{\alpha \beta}^A \hP_{\lambda \nu}^B )+\phi^K_{,\beta} \phi^I_{,\alpha}\phi^L_{,\lambda} \phi^J_{,\nu}\hat{R}_{LIJK}
\eea
where
\be
\Pi^I_{\mu \nu} =\nabla_\mu \nabla_\nu \phi^I+\hat{\Gamma}^I_{AB} \nabla_\mu \phi^A  \nabla_{\nu} \phi^B\,.
\ee
From this, one easily finds the explicit expressions for the induced Ricci tensor and Ricci scalar (we use the convention that for any rank-2 tensor $Y_{\mu \nu}$, $[Y] \equiv g^{\mu \nu} Y_{\mu \nu}$):
\bea
R_{\alpha \beta}[\tg]&=&A R_{\alpha \beta}[g]-\A_{IJ}  \nabla^{\nu} \phi^I  \nabla^{\lambda} \phi^J R_{\lambda \alpha \nu \beta}[g] \nn \\
&+& \a \H_{AB} \left(  [\hP^A] \hP^B_{\alpha \beta}  -(\hP^A \cdot \hP^B)_{\alpha \beta} \right) \nn \\
&+&\A_{IJ} \H_{AB} \left( (\partial \phi^I \cdot \hP^A)_\alpha (\partial \phi^J \cdot  \hP^B)_\beta -(\partial \phi^I \cdot  \hP^A  \cdot \partial \phi^J)  \hP^B_{\alpha \beta}  \right) \nn \\
&-&2(AX^{LJ}+2 X^{LM} X^{JN}\A_{MN})\hat{R}_{LIJK} \nabla_\beta \phi^K \nabla_{\alpha} \phi^I
\eea
and
\bea
R[\tg]&=&\a^2 R[g]-2\a \A_{IJ} (\partial \phi^I  \cdot R[g]  \cdot \partial \phi^J)+\A_{IJ} \A_{ML}  \nabla^{\lambda} \phi^I \nabla^{\nu} \phi^J  \nabla^{\mu} \phi^M \nabla^{\kappa} \phi^L R_{\lambda \mu \nu \kappa}[g]  \nn \\
 &+&\a^2\, \H_{AB} \,\left( [ \hP^A] [ \hP^B] -  [\hP^A \cdot \hP^B]\right)  \nn \\
 & -&2 \a \, \A_{CD} \, \H_{AB} \,\left(  [\hP^A] (\partial \phi^C \cdot \hP^B  \cdot \partial \phi^D)  -(\partial \phi^C \cdot \hP^A  \cdot \hP^B \cdot \partial \phi^D)\right)  \nn \\
 &+&(\A_{IJ} \A_{KL}-\A_{IL} \A_{KJ}) \H_{BE} \, (\partial \phi^I \cdot \hP^B  \cdot \partial \phi^J)   (\partial \phi^K \cdot \hP^E  \cdot \partial \phi^L)   \nn \\
 &+&4(AX^{LJ}+2 X^{LM} X^{JN}\A_{MN})(AX^{IK}+2 X^{IC} X^{KD}\A_{CD})\hat{R}_{LIJK}\,.
\eea
\\
Eventually, we have thus obtained the explicit expression
\bea
S_{{\rm ind}}=\int \dn{4}{x}\, \sqrt{-g}   \frac{\tM^2}{2} \left( \L_a +  \L_b+ \L_c+ \L_d+ \L_e+ \L_f +\L_g +\L_h   \right) 
\label{explicit-action}
\eea
where
\bea
\L_a &=&  \a^2 \sqrt{\cal D} R[g] \\
\L_b&=& -2  \a \sqrt{\cal D}  \A_{IJ} (\partial \phi^I  \cdot R[g]  \cdot \partial \phi^J)  \\
\L_c &=& \a^2  \sqrt{\cal D} \, \H_{AB} \,\left( [ \hP^A] [ \hP^B] -  [\hP^A \cdot \hP^B]\right) \\
\L_d &=&-  2 \a \sqrt{\cal D}  \, \A_{CD} \, \H_{AB} \,\left(  [\hP^A] (\partial \phi^C \cdot \hP^B  \cdot \partial \phi^D)  -(\partial \phi^C \cdot \hP^A  \cdot \hP^B \cdot \partial \phi^D)\right)  \\
\L_e&=& \sqrt{\cal D} (\A_{IJ} \A_{KL}-\A_{IL} \A_{KJ}) \H_{BE} \, (\partial \phi^I \cdot \hP^B  \cdot \partial \phi^J)   (\partial \phi^K \cdot \hP^E  \cdot \partial \phi^L)  \\
\L_f&=&\sqrt{\cal D} \A_{IJ} \A_{ML}  \nabla^{\lambda} \phi^I \nabla^{\nu} \phi^J  \nabla^{\mu} \phi^M \nabla^{\kappa} \phi^L R_{\lambda \mu \nu \kappa}[g]\\
\L_g&=& 4 \sqrt{\cal D} (AX^{LJ}+2 X^{LM} X^{JN}\A_{MN})(AX^{IK}+2 X^{IC} X^{KD}\A_{CD})\hat{R}_{LIJK} \\
\L_h&=&-\frac{3 \sqrt{{\cal D} }}{4}  \frac{f_{,I} f_{,J}}{f^2} \left( A X^{IJ}+2 \A_{KL} X^{KI} X^{LJ} \right)
\eea
and $\L_e, \L_f$ and $\L_g$ represent purely multifield effects. One should also note that perturbative calculations hint at a fully non-linear equality between $\A_{AB}$ and $H_{AB}$ although we were not able to prove it in the general case.

\section{Solutions to the constraint equations}
\label{Linear}

In this section, we give details about the constraints equations \refeq{N-constraint} and \refeq{Ni-constraint}, their solutions in different gauges, and the calculation of the second-order action.

\begin{itemize}
\item  {\bf Single field case and uniform inflaton gauge}
\end{itemize}

In the single field case, we write the scalar field Lagrangian as $P(X,\phi)$ where $X=-\frac12 g^{\mu \nu} \partial_{\mu} \phi \partial_{\nu} \phi$ and work in the uniform inflaton gauge in which $h_{ij}=a(t)^2e^{2 \z}\delta_{ij}$ and the field takes its unperturbed value. The linearised momentum constraint \refeq{Ni-constraint} then gives $V_i=0$ and
\beq
  \label{deltaN}
  \deltaN_{{\rm uni}} = \frac{\dot \z}{H} \X  \, ,
\eeq
where $\X$ was defined in Eq. \refeq{def-kappa}. The linearised energy constraint \refeq{N-constraint} then yields 
\beq
\beta_{{\rm uni}}=-\frac{\z}{aH} \X+a \chi 
\label{beta}
\eeq
where
\be
\partial^2 \chi = \frac{\M^2}{\M^2+\frac{\tM^2}{\c^3}}  \dot \z\left[ \frac{\dot \phi^2}{2 H^2 \M^2}\frac{\X \PX}{c_{{\rm k-inf}}^2}+  3(1-\X) +\frac{3 \tM^2}{\M^2 \c^3}\left(1 -\frac{(3-\c^2)\X}{2 \c^2}  \right) \right]
\ee
and $1/c_{{\rm k-inf}}^2-1\equiv  2XP_{,XX}/\PX$. Expanding the action to quadratic order in $\z$ by substituting\footnote{In the ADM formalism, it is sufficient to use the  perturbed lapse and shift up to first order as their second-order parts cancel out in the action \cite{malda,Chen:2006nt}.} the expression (\ref{deltaN}) for $\deltaN$ (as usual, the explicit expression of $\beta$ is not needed for the second-order action), one obtains
\bea
S_{(2)}&=& \int {\rm d}t\, \dn{3}{x} \,a^3 \M^2 \left[ 
\dot \z^2 \left(  \frac{\dot \phi^2}{2 H^2 \M^2}\frac{\X^2 \PX}{c_{{\rm k-inf}}^2}+  3(1-\X^2) +\frac{3 \tM^2}{\M^2 \c}\left(1 -\frac{(3-\c^2)\X^2}{2 \c^4}  \right)  \right)
\right.
 \cr
  &&   \hspace*{-1.5em} 
  \left.
 +\frac{(\partial \z)^2}{a^2} \left( 1+\frac{\tM^2 \c}{\M^2}-(1+\epsilon)\left(1+\frac{\tM^2}{\M^2 \c}\right)  \X   -\frac{1}{H}  \frac{{\rm d}}
{{\rm d}t} \left[   \left(1+\frac{\tM^2}{\M^2 \c}\right)  \X    \right]    \right)
    \right]\,, 
\label{2d-order-action}
\eea
where we have used the two background equations 
\be
 \left( \M^2+\frac{\tM^2}{\c^3} \right) 3H^2 +P-\PX \dot \phi^2=0
 \label{Energy0}
\ee 
and 
\be
\left( \M^2+\frac{\tM^2}{\c} \right) \left( 2 \frac{\ddot a}{a}+H^2  \right)-2H^2 \frac{\tM^2}{\c} \frac{\dot \c}{H \c}=-P\,,
\ee
in particular to show that the terms in $\zeta^2$ identically vanish. In our case of interest for which $P=-1/f (\sqrt{1-2f X}-1)-V$, this yields the result Eqs. \refeq{2d-order-action-2}-\refeq{B}.

\begin{itemize}
\item  {\bf Multifield case and flat gauge}
\end{itemize}

In the multifield case and in the flat gauge in which $h_{ij}=a(t)^2 \delta_{ij}$\,, the linearised momentum constraint \refeq{Ni-constraint} gives $V_i=0$ and
\bea
2H \M^2 \left(1+\frac{\tM^2}{\c^3 \M^2} \right) \deltaN_{\rm flat}=\left(1-3 \alpha \right) \frac{1}{\c}\dot \phi_I Q^I+ \frac{2 \alpha}{\c H} \dot \phi_I \dot Q^I 
\label{momentum}
\eea
while the linearised energy constraint \refeq{N-constraint} yields 
\bea
&&2H \left[ \left( \M^2 + \frac{\tM^2}{\c^3} \right) \frac{\partial^2 \beta_{\rm flat}}{a^2}  +\frac{f \tM^2}{\c^3}  \dot \phi_I   \frac{\partial^2 Q^I}{a^2}    \right] +\left(1-9 \q\right) \frac{\dot \phi_I \dot Q^I}{\c^3}+V_I Q^I  \nn \\
&+&  \deltaN_{\rm flat} \left[  -\frac{\sd^2}{\c^3} +6 H^2 \M^2 \left( 1+\left( \frac{3}{\c^2}-1 \right) \frac{\tM^2}{2\c^3 \M^2}   \right)   \right] =0\,.
\label{energy}
\eea
Note that, using the background equations of motion, one can check that the two above solutions agree, in the single field case, with the results obtained respectively from Eqs. \refeq{deltaN} and \refeq{beta} by making the explicit gauge transformation between the uniform inflaton gauge and the flat gauge:
\be
\deltaN_{\rm flat}=- \frac{1}{H} \frac{{\rm d}} {{\rm d}t} \left[ \frac{H}{\dot \phi}  Q    \right] (\X-1)+ \eps \frac{H}{\dot \phi}  Q 
\ee
and
\bea
\partial^2 \beta_{\rm flat}&=&\frac{ \partial^2 Q}{a\dot{\phi}}(\X-1) \nn \\
&-& a \frac{\M^2}{\M^2+\frac{\tM^2}{\c^3}} \left[ \frac{\dot \phi^2}{2 H^2 \M^2}\frac{\X \PX}{\cs^2}+  3(1-\X) +\frac{3 \tM^2}{\M^2 \c^3}\left(1 -\frac{(3-\c^2)\X}{2 \c^2}  \right) \right] \frac{{\rm d}} {{\rm d}t} \left[ \frac{H}{\dot \phi}  Q    \right]  \,. \nn
\eea
Eventually, let us give the form of the "anisotropy constraint", \textit{i.e.} the traceless part of the linearised gravitationnal $i-j$ equation of motion:
\bea
 &&\left(  \M^2+\frac{\tM^2}{\c}\right) \left( \deltaN_{\rm flat}+ \dot{\beta}_{\rm flat}-H \beta_{\rm flat}   \right) +2H \beta_{\rm flat} \left(\M^2+\frac{\tM^2}{\c}(1-s/2) \right) \nn \\
 &+&\frac{f  H \tM^2 \sd}{\c}\left(1-\frac{s}{1-\c^2} \right) e_{\s I} Q^I+\frac{f \tM^2 \sd}{\c}\dot{e}_{\s I}Q^I =0\,.
\eea
Given that the expressions in the flat gauge of two gauge-invariant Bardeen potentials are given respectively by $\Phi= \deltaN_{\rm flat} +\dot{\beta}_{\rm flat}$ and $\Psi=-H \beta_{\rm flat}$, one deduces from the above equation that they differ in our model, even without fluctuations of the scalar fields:
\bea
\left(  \M^2+\frac{\tM^2}{\c}\right) \Phi -\left(  \M^2+\frac{\tM^2}{\c}(1-s) \right) \Psi&=&-\frac{f  H \tM^2 \sd}{\c}\left(1-\frac{s}{1-\c^2} \right) e_{\s I} Q^I-\frac{f \tM^2 \sd}{\c}\dot{e}_{\s I}Q^I \,. \nn
\eea

\section{Calculation of the third-order action}
\label{Perturbations}

In this section, we give details about the calculation of the induced gravity action at third-order in the scalar perturbations, as needed in section \refeq{3rd}. We work at leading order in the slow-varying approximation and neglect the metric perturbations (although it is straightfoward to include them), which is consistent in the regime $\X \simeq 1$. The expressions of the various quantities needed when starting from the ADM expression of the induced gravity action (in the second line of Eq. \refeq{action-ADM-multifield}) are as follows (the field space indices are contracted with the trivial metric $G_{IJ}=\delta_{IJ}$):
\bea
\th_{ij}&=&a^2 \delta_{ij}+ f \partial_i Q^I \partial_j Q_I \equiv a^2 \delta_{ij}+ \th_{ij}^{(2)} \\
 \tN_i &=&  \tBone_{,i}+\tB_i^{(2)} \, 
\eea
with
 \bea
 \tBone &=&  f \sd \Qsi \\
 \tB_i^{(2)}&=& f \dQ_I \partial_i Q^I  
 \eea
With the notations
\bea
\tpone &=&-\frac{f \dot \phi_I \dot Q^I}{\c^2} \\
\tptwo&=& -\frac{f}{2\c^2} \dQ_I \dQ^I 
\eea
we also have
\bea
\frac{\c}{\tN}&=&1-\tpone -\tptwo+\frac32 (\tpone)^2-\frac{ (\partial \tBone)^2}{2a^2 \c^2} \nn \\
&-& \frac52 (\tpone)^3+3 \tpone \tptwo +\frac32 \tpone  \frac{(\partial \tBone)^2}{\c^2 a^2}- \frac{\partial^i \tBone \tB_i^{(2)}}{a^2 \c^2} +\CO(Q^4) \\
\sqrt{\th}&=&a^3 \left(1+ \frac{f}{a^2} \pQ^I \pQ_I \right) +\CO(Q^4) \\
\tilde{\Gamma}^i_{jk}&=&\frac{f}{a^2} \partial^i Q^I \partial_{j}  \partial_{k} Q_I +\CO(Q^3) \\
\tR^{(3)}&=&  \frac{f}{a^4} \left(  \partial^2 Q^I \partial^2 Q_I - \partial_i \partial_j Q^I \partial^i \partial^j Q_I \right)+\CO(Q^4) \\
\tE_{ij}&=& a^2 H\delta_{ij} - \partial_i \partial_j \tBone+\frac{1}{2}\dot{\th}_{ij}^{(2)}- \partial_{(i} \tB_{j)}^{(2)}+\frac{f}{a} \partial_i \partial_j Q_I \partial^k Q^I \partial_k \tBone + \CO(Q^4) \\
\th^{ij}&=& \frac{\delta^{ij}}{a^2}-\frac{f}{a^4} \partial^i Q^I \partial^j Q_I+\CO(Q^4) 
\eea
and
\bea
\tE_{ij} \tE^{ij}-\tE^2 &=& -6H^2+ 4H \frac{\partial^2 \tBone}{a^2} -\frac{2H}{a^2 }   \Tdot{ \delta^{ij} \left( {\tilde h}^{(2)}_{ij}   \right)} 
+4H  \frac{\partial^i \tB_i^{(2)}}{a^2} \nn \\
&+&\frac{1}{a^4} \left(\partial^i \partial^j \tBone \partial_i \partial_j \tBone -(\partial^2 \tBone)^2  \right)-4a^2 H^2 \delta_{ij}\th^{ij (2)} \nn \\
&-& \frac{4Hf^2 \sd}{a^4} \pQ^I \pQsi \partial^2 Q_I - \frac{1}{a^4} \left( \partial^i \partial^j \tBone-\delta^{ij} \partial^2 \tBone \right) \left(  \dot{h}_{ij}^{(2)} -2   \partial_{(i} \tB_{j)}^{(2)}\right) \nn \\
&-& \frac{2 H f^2 \sd}{a^4} \left( \partial^2 \Qsi \pQ^I \pQ_I +\partial_i \partial_j \Qsi  \partial^i Q^I \partial^j Q_I \right) +\CO(Q^4)\,.
\label{EE}
\eea
We then have all the building blocks to calculate the second and third-order action in this regime. For the former, this straightforwardly gives the result Eq. \refeq{S2flat}. More work is required for the latter. In particular, to deal efficiently with the term $- \frac{1}{a^4} \left( \partial^i \partial^j \tBone-\delta^{ij} \partial^2 \tBone \right) \left( \dot{h}_{ij}^{(2)} -2   \partial_{(i} \tB_{j)}^{(2)}\right) $ appearing in Eq. \refeq{EE}, it is useful to separate the scalar and vector contributions (in the sense of cosmological perturbations) from $\tB_i^{(2)}$ by writing
\be
\tB_i^{(2)}=f \left( \partial_i S + W_i \right)
\ee
with
\bea
S &=& (\partial^2)^{-1} \left(  \pdQ^I \pQ_I +\dQ^I \partial^2 Q_I \right)
\eea
and $\partial^i W_i=0$. One then obtains
\bea
&&- \frac{1}{a^4} \left( \partial^i \partial^j \tBone-\delta^{ij} \partial^2 \tBone \right) \left(  \dot{h}_{ij}^{(2)} -2   \partial_{(i} \tB_{j)}^{(2)}\right)  \nn \\
&& \hspace*{+0.5em} =\frac{2 f^2 \sd}{a^4} \left( \partial^2 \Qsi \pdQ^I \pQ_I -\partial^i \partial^j \Qsi \partial_i \dQ^I \partial_j Q_I +\partial^i \partial^j \Qsi \partial_i \partial_j S -(\partial^2  \Qsi) \partial^2 S \right)\,.
\eea
Integrating by parts to make appear $\partial^2 S$ and using Eq. \refeq{useful}, one finds
\bea
&& \int {\rm d}t \,\dn{3}{x}\, \frac{1}{a} \, \left(  \partial^2 \Qsi \pdQ^I \pQ_I -\partial^i \partial^j \Qsi \partial_i \dQ^I \partial_j Q_I +\partial^i \partial^j \Qsi \partial_i \partial_j S -(\partial^2  \Qsi) \partial^2 S \right)  \nn \\
&& \hspace*{+0.5em} =  \int {\rm d}t \,\dn{3}{x}\, \frac{1}{a} \, \left(    \frac{H}{2} \pQsi^2 \LapQsi + \doneperppuipuj \pdipdjQsi -  \doneLaponeperp \LapQsi  \right)\,.
\eea 
Together with a few non-trivial integrations by part listed in appendix \refeq{IPP}, and especially Eq. \refeq{precise}, one then obtains the result Eq. \refeq{S3ind}.

\section{Focusing on the induced gravity}
\label{Induced gravity}

We have seen in our discussion of the background evolution in section \ref{Background} that the induced gravity should be subdominant compared to standard gravity -- in the sense that $\tM^2 \ll \c^3 \M^2$ and hence $\X \simeq 1$ -- in order to achieve a phase of quasi de-Sitter inflation in the relativistic regime $\c^2 \ll 1$. Despite this, it is interesting to understand the opposite limit in which the induced gravity is dominant $\X \simeq \c^2$. In this appendix, we therefore consider the model defined by the action \refeq{action-f=cst} in the formal limit in which $\M \to 0$, thereby focusing on the effect of the induced gravity only. We do not make a comprehensive study of this model bur rather try to make the links with some of the general results in the main part of this paper.\\

In the main text, we argued that the single field second-order scalar action in terms of $\z$ Eq. \refeq{S2-MP=0} can be recovered by going to the Einstein frame in which $\tg_{\mu \nu}$ is treated as a cosmological metric and applying the standard results of $k$-inflation \cite{Garriga:1999vw}. Indeed, in the uniform inflaton gauge -- whose condition $\delta \phi=0$ reads the same in both frames -- the spatial parts of the metrics $g$ and $\tg$ coincide (see Eq. \refeq{h}). Hence, the scalar degree of freedom in the Einstein frame ${\tilde \z}$, defined such that $\th_{ij}=a^2(t) e^{2 {\tilde \z}}$, actually coincides with $\z$, such that $h_{ij}=a^2(t) e^{2\z}$. As for the lapse functions $N$ and $\tN$ and the shift vectors $N_i$ and $\tN_i$, they appear in both frames as auxiliary variables, and extremizing the action with one set or another leads to the same constraints. Therefore, for the purpose of determining the scalar action in terms of $\z$ at any perturbative order, treating $g$ or $\tg$ as the metric of reference represents a mere change of variables in the intermediate calculations.\\

The situation is more subtle in the multifield situation. Indeed, the standard variables to use then are the field fluctuations in the spatially flat gauge. However, the two spatial metrics being related by $\th_{ij}=h_{ij} +f G_{IJ} \partial_i \phi^I \partial_j \phi^J$ (Eq. \refeq{h}), the spatially flat gauge for the cosmological metric does not coincide with the spatially flat gauge for the induced metric, because of the spatial gradients of the scalar fields. However, the two gauges coincide at linear order in the fluctuations. Hence, expressing the brane action as $S_\matter =  \int d^4 x \sqrt{-\tg} F(X_{\tg}^{IJ},\phi)$ where $X_{\tg}^{IJ} \equiv -\frac{1}{2}\tg^{\mu \nu}\partial_{\mu} \phi^I \partial_{\nu} \phi^J$, one can readily apply the linear analysis of Ref. \cite{Langlois:2008qf} for such scalar field Lagrangians without making any change of variables. One then easily recovers the $\M \to 0$ limit of the result Eq. \refeq{S_v}. In particular, note that the speed of propagations of scalar perturbations with respect to cosmic time is then unity.\\

From the analysis of the linear cosmological perturbations, one deduces that the latter are not ghosts if and only if $\teps > 0$ or equivalently, with
\be
\teps=\frac{3}{2} \frac{f \sd^2}{\c} \frac{1-fV}{1-\c +\c fV}\,,
\label{Hdot-induced}
\ee
$1-f V >0  \Leftrightarrow  1-3 \q >0 $. Given that Eq. \refeq{Hdot-induced} is equivalent to
\be
\ddot \sigma +3H \dot \s +\frac{\c^2}{1-fV}V_{,\s}=0\,,
\ee
where $V_{,\s} \equiv e_\s^I V_{,I}$, one sees that the fields tend to climb up the potential precisely in the regime in which the fluctuations are ghosts. Note also that the deceleration parameter $\eps$ is given by $\teps-s$, so that one can imagine, if $s \equiv {\dot c}_\D/(H \c)$ is sufficiently large, a ghost-free regime ($\teps > 0$) violating the null energy condition $\eps < 0$ and in which the speed of propagation of all scalar fluctuations is unity (this should be contrasted with the results in Ref. \cite{Dubovsky:2005xd} for example). Finally, note that if $s$ can be neglected, $\eps \simeq \teps$ can be small in the ghost-free relativistic regime only if $1-fV \ll \c$, in other words, at the expense of having very small kinetic terms for the scalar fields and hence at the risk of being in a strongly coupled regime.

\section{ Effects of the warping}
\label{warping}

In this section we discuss some of the consequences of considering a general non-constant warp factor $h$ in our starting point action Eq. \refeq{action}. 
Using the general gravitational \refeq{metric-eoms} and fields \refeq{scalar-eoms} equations of motion, and employing the same computational trick as in section \ref{Background}, we easily determine the modified Friedmann equations
\bea
\label{Friedmann00}
3 H^2 \M^2+ 3\hat{H}^2 \frac{ \tM^2}{c_D^3}
 =\rho_{\matter}=
V+ \frac{1}{f} \left(\frac{1}{c_D}-1\right)
\eea
and
\bea
-\M^2 \dot{H} -\frac{\tM^2}{\c} \dot{\hat{H}}
&=& \frac{\sd^2}{2 c_D}
- \frac{\tM^2}{c_D}
\left[ \frac32 \left(\frac{1}{c_D^2} -1\right)
\hat{H}^2-\frac{\c}{h^{\frac14}}
 \Tdot{\left(\frac{h^{\frac14}}{\c}\right)} 
\hat{H}\right]
 \label{Friedmannij}
\eea
where $\hat{H}$, defined by
\be
\hat{H} \equiv H-\frac{\dot{f}}{4 f}\,,
\ee
differs from the Hubble parameter due to the non-trivial warping. As for the equations of motion for the scalar fields \refeq{scalar-eoms}, they read, in the background:
\bea
&&\frac{1}{\c^2}\left(3 \hat{H}^2 \frac{\tM^2}{\c^2}
- \frac{1}{f}
\right)
\left(\mathcal{D}_t \dot{\phi}^I +
\frac{\sd^2}{2f} (e^I_\s e^J_\s    - \perp^{IJ}) f_{,J}
-\frac14 \frac{f^{,I}}{f^2} c_D^2\right)\nn\\
&&+\left[ \frac{\tM^2}{\c^2}
\left(2 \dot{\hat{H}}+3 \hat{H}^2
+2 \left(\frac{\c}{h^{\frac14}}\right)
 \Tdot{\left(\frac{h^{\frac14}}{\c}\right)} 
\hat{H}\right)-\frac{1}{f}\right]
\left(3 H \dot{\phi}^I -
\frac{3}{4}  \frac{f^{,I}}{f^2} \right)\nn\\
&&
-\frac{1}{\c f} (c_D^2 e^I_\s e^J_\s  + \perp^{IJ})
\left(V_{,J} + \frac{f_{,J}}{f^2}\right)=0\,,
\label{eom_field}
\eea
where, like in the main text, $e^I_{\s} \equiv \dot \phi^I/\sd$ is the unit vector (with respect to the field space metric) pointing along the background trajectory in field space, we have introduced the projector on the subspace orthogonal to it
\bea
\perp_{IJ} &\equiv& G_{IJ}-e_{\s I} e_{\s J}
\eea
and 
\be
{\cal D}_t \dot \phi^I \equiv \ddot \phi^I+\Gamma^I_{JK} \dot \phi^J \dot \phi^K
\ee
represents the acceleration vector of the fields in curved coordinates (in field space) \cite{GrootNibbelink:2000vx,Bartjan-linear}. Because of the non-trivial warping, second-order derivatives of the fields appear not only through $\dot{c}_\D$ in Eq. \refeq{Friedmannij} and through ${\cal D}_t \dot \phi^I$ in Eq. \refeq{eom_field} but also in $\dot{{\hat H}}$, which contains
\be
\ddot f=f_{,I} \ddot \phi^I +f_{,IJ} \dot \phi^I \dot \phi^J \,.
\ee 
Hence, should one wish to express separately and explicitly the second-order derivatives of the scale factor and of the fields, one would first need to solve a system of three linear equations for the three unknows $\dot H$, $\sdd \equiv e_{\s I} {\cal D}_t \dot \phi^I $ and $\perp_{I}^{J} f_{,J}  {\cal D}_t \dot \phi^I$: Eq. \refeq{Friedmannij} and the contractions of Eq. \refeq{eom_field} with $e_{\s I}$ and $ \perp_I^J f_{,J}$. This can easily be carried out but the explicit solutions are not particularly illuminating and we do not reproduce them. On the other hand, in the quasi de-Sitter slow-varying relativistic regime described in section \refeq{Background}, the equations of motion for the fields take the simple approximate form:
\bea
&&3H \sd(1-3 \q)+\c V_{,\sigma} \simeq 0\,,
\label{eom_ad_leading}\\
&&\perp_{I J} {\cal D}_t e_{\s}^J \simeq-\frac{\perp_I^J}{\sd} \left(
\frac{\c}{1-3\q} V_{,J}  - \frac{f_{,J}}{2f^2}  \right)\,,
\label{eom_en_leading}
\eea
where the induced gravity effects appear in the same combination $1-3\q$ as in the gravitational equation \refeq{eps-approx}. Eventually, let us note that the same complexity as described above holds of course at the level of the perturbations: due the non-zero gradients of the warp factor along the entropic directions in field space, second-order time derivatives of the adiabatic and entropic perturbations do not decouple in the equations of motion, contrary to Eqs. \refeq{eq_v_sigma} and \refeq{eq_v_s}. Hence, although there is no conceptual obstacles to deriving the exact second-order action in that case, the result would be of little practical analytical use.

\section{Useful integrations by part}
\label{IPP}

We collect here a number of non completely trivial integrations by part needed in the calculation of the third-order action in appendix \ref{Perturbations}.

\bea
\int {\rm d}t \,\dn{3}{x}\, \frac{1}{a} \,  \left( \dQsi   \partial_i \partial_j \Qsi \partial^i \partial^j \Qsi  - \dQsi (\LapQsi)^2 \right)  &=&  \int {\rm d}t \,\dn{3}{x}\, \frac{1}{a}  \left( \frac{1}{2}H (\pQsi)^2 \LapQsi \right)    \\
\int {\rm d}t \,\dn{3}{x}\, \frac{1}{a} \, \left( \LapQsi \pQsi \pdQsi - \partial_i \Qsi \partial_j \dQsi \partial^i \partial^j \Qsi  \right)  &=&  \int {\rm d}t \,\dn{3}{x}\, \frac{1}{a}  \left( \frac12 H (\pQsi)^2 \LapQsi    \right) 
\label{useful}
\eea

\be
\int {\rm d}t \,\dn{3}{x}\, \frac{1}{a} \, \left( \dQsi   \partial_i \partial_j \Qs \partial^i \partial^j \Qs  - \dQsi (\LapQs)^2\right) =  \int {\rm d}t \,\dn{3}{x}\, \frac{1}{a}  \left( \LapQs \pdQsi \pQs +\frac12  \gradperp \partial^2 \dQsi   \right)  
\ee

\be
\int {\rm d}t \,\dn{3}{x}\, \frac{1}{a} \,  \partial_i \partial_j \Qsi \partial^i \Qs \partial^j \Qs  =  \int {\rm d}t \,\dn{3}{x}\, \frac{1}{a}  \left(  -\LapQs \pQsi \pQs+\frac12 \LapQsi \gradperp \right)  
\ee

\bea
&&\int {\rm d}t \,\dn{3}{x}\, \frac{1}{a}  \left(
2 \doneperppuipuj \pdipdjQsi
-2  \doneLaponeperp \LapQsi
+\frac{1}{2} \gradperp \partial^2 \dQsi  
+ \pdQsi \poneLaponeperp
\right) \nn \\
 && \hspace*{+0.5em} =\int {\rm d}t \,\dn{3}{x}\, \frac{H}{a}  \left(
 \pQsi \pQs \LapQs+\frac12 \gradperp \LapQsi
\right) 
\label{precise}
\eea

\bibliography{Biblio}

\end{document}